\documentclass{aa}  
\usepackage[switch]{lineno} 
\usepackage[english]{babel}
\usepackage{natbib}
\usepackage{appendix}
\usepackage{txfonts}
\usepackage{amsmath}
\usepackage{graphicx}
\usepackage[colorlinks=true, allcolors=blue]{hyperref}
\usepackage{xcolor}
\newcommand{\opm}[1] {{#1}}

\newcommand{\HI}{H\,{\small I}}

\newcommand{\kms}{$\,$km$\,$s$^{-1}$}
\newcommand{\WHz}{$\,$W$\,$Hz$^{-1}$}

\newcommand{\mJybeam}{mJy beam$^{-1}$}

\newcommand{\msun}{{$M_\odot$}}
\newcommand{\msunyr}{{$M_\odot$ yr$^{-1}$}}

\newcommand{\tspin}{$T_{\rm spin}$}

\def\OIII{{[\ion{O}{III}]}}
\def\OII{{[\ion{O}{II}]}}
\def\HI{\ion{H}{I}}

\def\emph#1{{\sl #1}}
\newcommand{\ltsima} {$\; \buildrel < \over \sim \;$}
\newcommand{\gtsima} {$\; \buildrel > \over \sim \;$}
\newcommand{\lta} {\lower.5ex\hbox{\ltsima}}
\newcommand{\gta} {\lower.5ex\hbox{\gtsima}}

\begin{document} 

   \title{A deep MeerKAT view of associated \HI\ absorption in radio AGNs at intermediate redshift:  Role of absorber geometry and conditions of the gas } 

\authorrunning{Morganti et al.}
\titlerunning{\HI\ absorption observations with MeerKAT of AGNs at intermediate redshift}

\author{Raffaella Morganti\inst{1,2}, Tom Oosterloo\inst{1,2}, Clive Tadhunter\inst{3}, Suma Murthy\inst{4,5}}

    \institute{ASTRON, the Netherlands Institute for Radio Astronomy, Oude Hoogeveensedijk 4, 7991 PD, Dwingeloo, The Netherlands. 
 \email{morganti@astron.nl}
    \and
    Kapteyn Astronomical Institute, University of Groningen, Postbus 800, 9700 AV Groningen, The Netherlands
    \and
    Department of Physics and Astronomy, University of Sheffield, Sheffield, S7 3RH, UK
    \and
    Joint ALMA Observatory, Alonso de C\'ordova 3107, Vitacura, Casilla 19001, Santiago de Chile, Chile
    \and
    Joint Institute for VLBI ERIC, Oude Hoogeveensedijk 4, 7991 PD Dwingeloo, The Netherlands. 
}
 \abstract
{ We present  MeerKAT observations searching for \HI\ absorption in a sample of 17 powerful ($L_{\rm 1.4GHz}> 10^{27}$ \WHz) radio sources at intermediate redshifts ($0.25<z<0.7$). 
The sample is well characterised at radio and optical wavelengths, allowing us  to connect the presence (or absence)  of \HI\ to the properties of the active galactic nucleus (AGN) and its host galaxy. The sample consists mostly of core-dominated sources and quasars. Half of the targets have a  UV luminosity below the limit of $L_{\rm UV} = 10^{23}$ \WHz\,, whereby at values   above this limit, the gas would be expected to be ionised by this radiation.  We obtained 15 spectra free (or almost free) of radio frequency interference, reaching extremely low optical depths ($\tau_{\rm peak} < 0.005$)  resulting in three new \HI\ absorption detections. Two are associated \HI\ absorptions, giving  a detection rate of such systems of $13\%\pm 7\%$.
Both are found in  a young radio source (PKS~1151--34 and PKS~1306--09), confirming the trend that this type of sources are more often detected in \HI\ compared to more evolved ones. The UV luminosity of both these sources is below  $10^{23}$ \WHz. 
Surprisingly, one of the detections (PKS~1151--34) is hosted by a quasar, suggesting that the radio lobes of this source are still embedded in the circumnuclear disc. In the second source (PKS~1306--09), the \HI\ is highly blueshifted and likely part of the jet-driven outflow earlier observed in the warm ionised gas. This represents a new addition to the group of young radio AGNs, where multi-phased outflows have been observed as predicted by numerical simulations.
A third detection is a 'local intervening' system, caused by a galaxy in the local environment of  PKS~0405--12 and located in front of the southern radio lobe of this source, about 100 kpc in projection from this quasar.  
More such cases are expected to show up in large, blind surveys and our results show the need for high spatial resolution and good ancillary data to separate associated from intervening absorption. Overall, the results indicate a variety of plausible situations, which resemble what is seen at low redshifts. For the associated absorption, a combination of evolutionary status of the radio sources, physical conditions, and geometry of the gas structure determine the detection rate of \HI\ absorption.
The data also show the excellent capabilities of MeerKAT for obtaining very low optical depth detections, revealing the presence of an otherwise missed group of absorptions.}

   \maketitle

\section{Introduction}
\label{sec:Introduction}

The presence and properties of the  cold gas ($T \ll10^4$K) in galaxies \opm{ play a crucial role in  tracing their evolution} and  star formation history  (e.g.\ \citealt{Saintonge22,Crain23}). Furthermore, this gas can be relevant for providing the fuel for turning a dormant super-massive black hole (SMBH) into an active galactic nucleus (AGN), as well as for tracing the effects of the energy released by the AGN. Thus, there are several reasons for studying the cold gas in galaxies hosting an AGN. Although the various phases of this gas (atomic neutral and molecular) are usually detected in emission,  absorption observations can provide a complementary way of tracing this gas (e.g.\ \citealt{Morganti24,Rose24}).

\HI\ 21-cm absorption observations have  been used for many years to trace both intervening gas (i.e.\  gas unrelated to the AGN host and observed along the line of sight; \citealt{Kanekar04}), as well as  gas associated with the AGN host (see \citealt{Morganti18,Morganti24} for overviews). Although  they differ in nature, they are both important for tracing the evolution of the presence and properties of \HI\ in and between galaxies. 

Here, we focus on associated \HI\ absorption which traces atomic hydrogen located mainly in the central circumnuclear regions of radio AGNs. The interpretation of absorption observations is complicated by the fact that only  gas in front of the radio continuum can be traced. In addition, due to the dependence of the optical depth on the spin temperature, there is a bias towards detecting the colder \HI. However, these observations do offer other advantages that can be exploited. In particular, using VLBI observations, absorption observations can trace \HI\ gas down to pc scales  and --- as long as the background continuum is strong enough --- it can be used to trace the \HI\ in objects with redshifts much higher than can typically be done using \HI\ in emission.
Here, we focus on \HI\ absorption observations of a sample of radio sources at intermediate redshifts ($0.26 < z < 0.67$).

According to  \cite{Aditya24}, about a thousand radio sources (mostly radio AGNs) have been observed to investigate their associated \HI\ absorption,  with the vast majority of the  detections being made at low redshift (i.e.\ $z<0.2$ or, equivalently, at $\nu > 1200 $ MHz), where the detection rate ranges between 10 and 40\%\ (see below). 
The  redshift limit of $z=0.2$ is mostly set by the limited capabilities of the receivers  available in radio telescopes to observe at lower frequencies. However, the
situation has changed in recent years, thanks to radio telescopes such as the uGMRT, MeerKAT, and ASKAP,  thanks to their capacity to conduct observations at frequencies well below 1200 MHz. For the study of \HI\ in radio AGNs,  observations using radio interferometers are more suitable than those done with single-dish telescopes because of their higher angular resolution and bandpass stability; the relevance of this will become clear in the present study. 

For low-redshift sources, as we now know thanks to recent absorption studies, \HI\ can be associated with all the key circumnuclear structures of AGNs (see e.g.\ \citealt{Morganti18} for a review).  The most prevalent \HI\ structures  are discs and rings. The gas in these  is not always kinematically relaxed, but can have high velocity dispersion and/or is undergoing fast outflows (see \citealt{Morganti03} and \citealt{Murthy24} and refs therein for examples). In addition, clouds of infalling gas are seen in some objects (\citealt{Gorkom89,Tremblay16,Maccagni18,Maccagni23} and refs therein), \opm{possibly related to   feeding  the SMBH}. More rarely, clouds of quiescent gas at larger distances from the centre, albeit still bound to the host galaxy, are observed (e.g.\  \citealt{Morganti09,Struve12}). 

The detection of \HI\ in absorption is also connected to the type of radio AGNs observed and this reflects on their detection rate.
For example, the group of young (or recently restarted) radio sources (with ages $< 10^6$ yr), which are identified by being compact and having a peaked or steep radio spectrum (i.e. CSS and GPS, \citealt{ODea21}), have the highest detection rate at both low and high redshift (see below for more details). 
This is likely due to the combination of two factors. The first is the (sub-)kpc size of the radio continuum of these sources, which roughly matches the scale of the gas discs and rings. The second is the fact that these young sources are typically in a gas-rich phase which also provides the fuel for the radio AGN.
Detection rates up to 40\% have been found for such sources (e.g.\ \citealt{Vermeulen03,Gupta06,Chandola11,Gereb15,Glowacki17}) and the \HI\ often shows disturbed kinematics (e.g.\  \citealt{Struve12,Schulz18,Schulz21,Murthy24}), suggesting an interaction between the jets and the ISM, as also predicted by the simulations \citep{Mukherjee18,Perucho24,Dutta24}.

Extended radio sources show a lower detection rate (roughly  10\%; \citealt{Morganti01,Chandola13}), but with large scatter due to the low number of objects observed. However, broad-line galaxies and BL~Lac objects are typically not detected. This suggests the \HI\ is mostly associated with nuclear tori and discs which would be, as predicted by unified schemes of AGNs \citep{Tadhunter08},    seen face-on in broad-line
objects and not, thus, in front of the radio continuum.

Unsurprisingly, the detection rate of \HI\ in absorption is lower than found for \HI\  emission (about 40\%; \citealt{Serra12}) for early-type galaxies (the typical hosts of radio AGNs); nevertheless, the results above show that \HI\ absorption can help trace the presence of \HI. In this respect, it is also interesting to explore  higher redshift objects that  emission studies cannot reach easily.

\section{Associated \HI\ absorption at higher redshift ($z>0.2$) }
\label{sec:overview}
 
Studies of associated \HI\ absorption at higher redshift tend to show a more complex view of the presence and properties of this gas, although the  number of observations is still relatively small (see  the reviews of \citealt{Dutta22} and \citealt{Morganti24}). 
In general, there is a tendency to find a lower detection rate ($\ll10$\%) of \HI\ absorption at $z>0.2$, which becomes even more pronounced for $z>1$ \citep{Curran08}.
There is no  agreement  on what is the cause of this trend. 
The physical conditions of the gas can be influenced by high levels of UV radiation, which can be expected to be present in the powerful sources typically observed at high $z$. \cite{Curran08} proposed a limit to the UV luminosity  of $L_{\rm UV}=10^{23}$ \WHz\ above which the \HI\ would be ionised, thereby reducing the detection rate \citep{Curran08,Curran13,Aditya16,Curran17,Aditya18a}. 
However, that other factors may also be relevant is suggested  by the fact that \HI\ absorption has been detected in some objects with a UV luminosity higher than the $10^{23}$ \WHz\ limit (e.g.\ \citealt{Aditya17,Aditya21}) and by the low detection rate even for samples of sources selected with $L_{\rm UV}<10^{23}$ \WHz\ \citep{Murthy22}.

\begin{figure}
\centering
\includegraphics[width=0.95\linewidth]{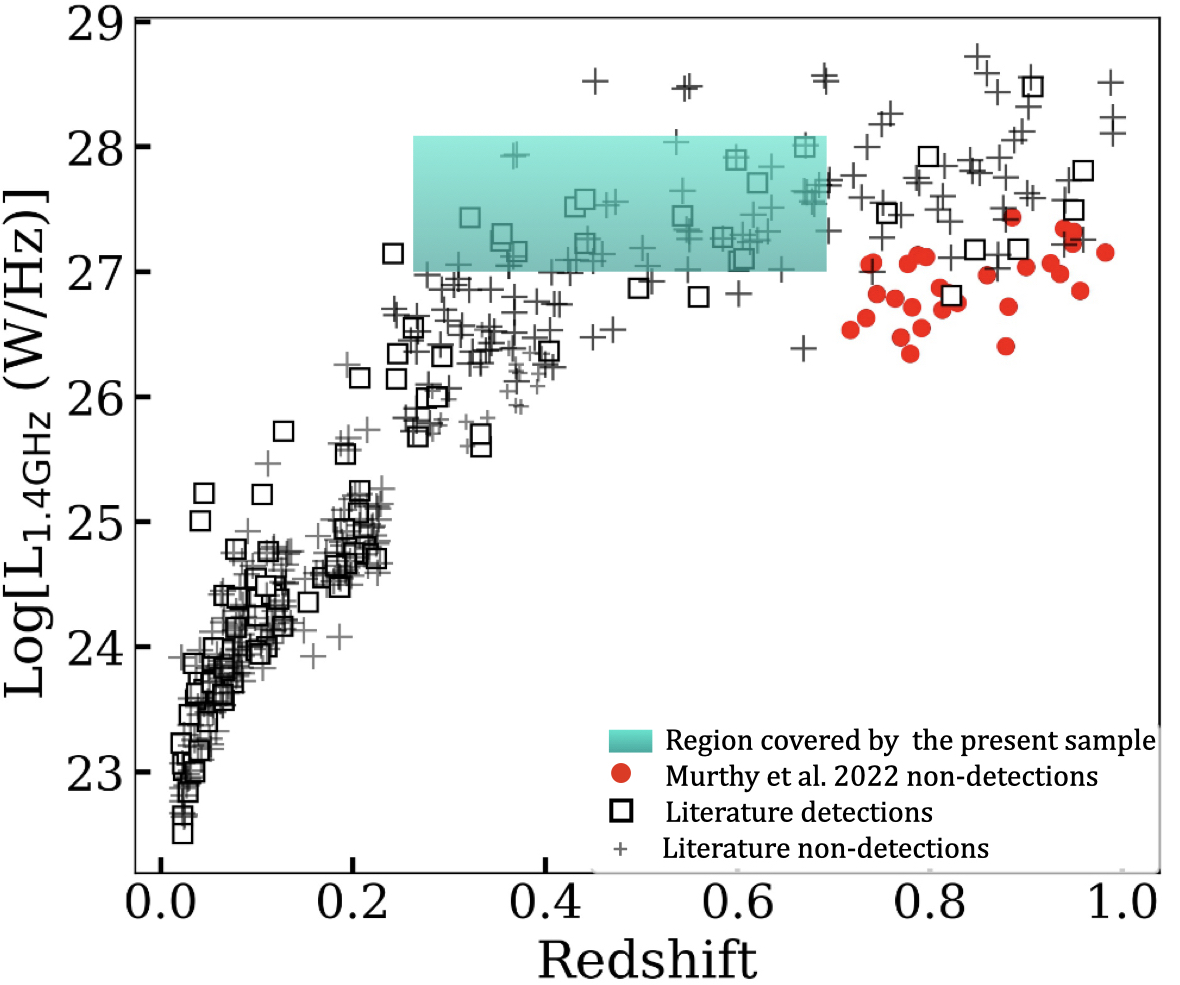}
\caption{\label{fig:Distribution-z} Distribution of the radio luminosities for sources in various samples available in literature, adapted from \cite{Murthy22}. Location of the present sample is indicated as the green area.}
\end{figure}

As an alternative, the detection rate may depend on the properties and the types of sources observed. As can be seen from Fig. \ref{fig:Distribution-z}, most of the AGNs observed in \HI\ absorption at redshift $z > 0.2$ (referred to as high-$z$ AGNs in this work) are different from those observed at low $z$; they typically have higher radio power ($ L_{\rm 1.4~GHz}>10^{26}$ \WHz) than the low-$z$ sources which mostly lie  below this limit (see Fig.\ \ref{fig:Distribution-z}). In addition, the high-$z$ samples have a higher fraction of quasars and flat-spectrum objects (i.e.\ most targets have been selected from flat spectrum samples). In such cases, the detection of  a circumnuclear \opm{disc of \HI\ can be hampered by orientation effects as is seen  at low $z$ (e.g.\ \citealt{Murthy21}) }.
The presence of a strong radio AGNs can also affect the detectability by the increase in the \tspin\ of the gas in the nuclear regions due to the strong radio flux \citep{Bahcall69} and the consequent decrease in the optical depth for a given column density.  

Interestingly, several high-$z$ studies find that compact, peaked spectrum sources tend to maintain their higher detection rate at high redshifts (see e.g.\ \citealt{Aditya18b}), although this finding is still based on  limited statistics. This appears to be further confirmed by the initial results of blind surveys such as FLASH \citep{Yoon25}. Thus, the conditions of the gas in the very centre  can can make it possible to detect the presence of atomic hydrogen even at high $z$.

In summary,  \HI\ absorption studies at $z > 0.2$ are giving mixed results about whether \HI\ is indeed  present less frequently at $z>0.2$ in radio AGNs (unlike what is seen in molecular gas; \citealt{Audibert22}).
Understanding the role of selection effects not only requires larger samples, but also a better characterisation of the targets.  
While the blind surveys that are now becoming available (e.g.\ FLASH; \citealt{Yoon25}) 
can give a view of the presence of \HI\ in a larger number of objects, there is still a need to investigate possible trends and biases and connect the detection rates and properties of the absorption. To do so, for example, we can use the  strength and morphology of the background continuum, the types of radio and optical AGNs, and other factors.

\begin{table*}
\caption{Observational parameters. }
\begin{center}
\small
\begin{tabular}{lccccr@{ $\times$ }lcccc} 
\hline\hline 
 Source  & &  $z$ & $\nu_{\rm obs}$ & $\Delta\nu$ & \multicolumn{2}{c}{Beam} & PA  &  $\sigma_{\rm cont}$ & $\sigma_{line}$     & Notes/Refs\\
        & &     & (GHz)                 &    (\kms)   & (" & ") & ($^\circ$)&   (\mJybeam) & (\mJybeam) &\\
\hline
 PKS 0159--11 & Q & 0.669 & 851.05 &  11.7 &  9.3 &   8.6  & \phantom{14}0.9 & 0.08 & 0.64   & 1,5 \\
 PKS 0252--71 & G & $0.56443\pm 0.00003^a$ & 907.02 & 11.0 &  13.3 &   6.3  & 144.1 & 0.14 & 1.74   &  2,3,5\\
 PKS 0403--13 & Q & 0.571       & 904.14 & 11.0 & 8.9 & 7.6 & 166.3 & 0.30 & 0.85 & 4,5\\
 PKS 0405--12 & Q & 0.574       & 902.41 & 11.0 & 9.3 & 8.3  & \phantom{1}40.2 & 0.37 & 0.68  & 5,4\\
 PKS 0637--75 & Q & 0.651       & 860.33 & 11.0 &17.4 & 6.2 & 103.9 & 0.11 & 1.07 & 6\\
 PKS 0842--75 & Q & 0.524       & 932.02 &  -- & \multicolumn{2}{c}{--} & -- & -- & --  & 6 (RFI) \\
 PKS 0859--25 & G & 0.305       & \llap{1}088.43 &   -- & 7.8 & 6.0 & 171.1 & 0.17 & -- & 5 (RFI) \\
 PKS 1136--13 & Q & 0.544       & 914.03 & 10.9 & 9.6 &  7.2 & 150.9 & 0.35& 0.91 & 5,7\\
 PKS 1151--34 & Q & $0.2579\pm 0.0001^a$  & \llap{1}129.09 & 6.9 & 9.5 & 5.8 & 141.2  & 0.57 & 1.86  & 2,3,5\\
 PKS 1306--09 & G & $0.46692 \pm 0.00007^a$ & 970.22 & 10.2 & 12.2 & 6.9  & 159.7 & 0.60 & 1.14  & 2,3,5\\
 PKS 1355--41 & Q & 0.313       & \llap{1}081.80 & 14.5 & 9.8 & 6.2 & 132.2 & 0.08 & 0.66  & 6\\
 PKS 1510--08 & Q & 0.361       & \llap{1}043.64 & 7.5 & 10.5 & 6.7 & 131.3 & 0.09& 0.69 & 5,8\\
 PKS 1602+01  & BLG & 0.462     & 971.55 & 10.2 & 11.5 & 8.3 & 116.4 & 0.10 & 0.97  & 5,9,10\\
 PKS 1954--388 & Q & 0.63\phantom{0} & 871.41 & 11.4 & 10.2 & 7.1 & \phantom{1}20.4 & 0.05 & 0.55 & 5\\
 PKS 2203--18  & Q & 0.618      & 877.87 & 11.3 & 9.2 & 7.5 &  170.2 & 0.24 & 1.19 & 5\\
 PKS 2243--123 & Q & 0.63\phantom{0} & 871.41 & 11.4 & 9.1 &  8.3 & 176.4 & 0.07 & 0.59  & 4,5,11\\
 PKS 2345--16  & Q & 0.576      & 901.27 & 11.0  & 10.1 & 8.1  & \phantom{1}48.1 & 0.14& 0.77  & 4,5\\
\hline
\end{tabular}
\tablefoot{The redshifts of most of the observed galaxies are taken from \citet{Tadhunter93}. $^{(a)}$ Redshifts are taken from \citet{Santoro20} and are based on the measurement of the stellar absorption lines associated with the host galaxy starlight. Note that  \citet{Santoro20} did not correct the redshifts to heliocentric. For PKS~1151--34 this correction for the redshift is +0.000083, while for PKS~1306--09 it is +0.000096. The beam sizes are those of the continuum images. References: (1) \cite{Reid99}; (2) \cite{Tzioumis02}; (3) \cite{Santoro20}; (4) \cite{Cheng20}; (5) \cite{Morganti93}; (6) \cite{Burgess06}; (7) \cite{Fernini14}; (8) \cite{Lister09}; (9) \cite{Morganti99};(10) \cite{Tadhunter02}; (11) \cite{Cheng23}}
\end{center}
\label{tab:obs}
\end{table*}

Here, we present the search for \HI\ absorption in a sample of powerful radio sources (see Fig.\ \ref{fig:Distribution-z}) extracted from the 2-Jy sample, which has been well studied in several wavebands  (see description below).  
We have taken advantage of the capabilities and the location of MeerKAT telescope to observe sources in the redshift range $0.2<z<0.7$. For the strong sources in the sample, we have been able to  reach very low optical depths of the \HI\ 21-cm absorption thanks to the stability of the MeerKAT's receiver system.

The paper is organised as follows. In Sect.\  \ref{sec:Targets}, we describe the selection of the targets with a brief recap of the 2-Jy sample. In Sect.\ \ref{sec:Observations}, we describe the new MeerKAT observations and the data reduction. In Sect.\  \ref{sec:Results}, we describe the general findings  as well as the properties of the detected targets. We discuss the possible implications of our findings in Sect.\ \ref{sec:Discussion}. All distances, spatial scales, and luminosities were calculated assuming a flat Universe with $\Omega_{\rm M} = 0.286$, $\Omega_{\rm vac} = 0.714$, and $H_\circ = 69.6$ \kms\ Mpc$^{-1}$.

The target sources represent a complete sub-set of the so-called 2-Jy sample including southern ($\delta < +10^\circ$) radio sources with high 2.7-GHz radio fluxes ($S_{\rm 2.7~GHz} > 2$ Jy; see \citealt{Tadhunter16} for an overview of the 2-Jy sample). 
Here, we focus on sources with redshift $0.25<z<0.7$ and the targets and their redshifts are listed in Table \ref{tab:obs}.
The sample has been well characterised in multiple bands, especially at radio (VLA/ATCA/VLBI; \cite{Morganti93,Morganti01,Tzioumis02,Morganti99} and optical wavelengths (NTT/VLT/Gemini; \citealt{Tadhunter93,Tadhunter98,Holt08,Ramos11}). 
Because of the redshift limit ($z<0.7)$, the presence of optical emission lines, including H$\alpha$+[\ion{N}{ii}],  could be observed for the sources in the sample to fully characterise the spectral properties. 
Part of the sample has also been observed with ALMA to investigate the properties of the molecular gas (in particular, focussing on CO(1-0); \citealt{Tadhunter24}), while the study of the cold ISM traced by Herschel is reported in \cite{Bernhard22} and \cite{Dicken23}.
However, some of these studies have excluded the flat spectrum sources which dominate the present sample (see below).

Because of a limited allocation of observing time (and on request of the MeerKAT time allocation committee), we have restricted the observed sample to sources with flux larger than 30~mJy (at 1.4~GHz)  of the core or the central component.
This resulted in 17 sources with $0.25<z<0.7$, with a strong preference for core dominated objects; of these sources, only 4 are radio galaxies, of which 1 is classified as a broad-line galaxy \citep{Tadhunter02}. The rest are quasars (i.e.\ with broad permitted emission lines seen in the optical spectrum; see  notes in \citealt{Tadhunter93}). In radio, a number of them are variable and have been  monitored at high spatial resolution and up to high energies (including $\gamma$-rays). In Table \ref{tab:obs}, we list some of the references to available radio observations. 

Four of the targets are classified as CSS and, with the exception of PKS~1306--09, a peak in the radio SED at low frequency is also observed \citep{Tzioumis02,Callingham17}.
Three of them (PKS~0252--71, PKS~1151--34 and PKS~1306--09) have been observed in radio with Very Long Baseline Interferometry by \cite{Tzioumis02} and were studied in detail in the optical using X-shooter by \cite{Santoro20}.
Radio observations of a fourth candidate (quasar PKS~2243--123) are presented in \cite{Cheng23}. As mentioned in the introduction, CSS sources are considered to be young radio sources where a newly born jet is in the process of expanding in the central kpc regions \citep{ODea21} 

\section{The targets}
\label{sec:Targets}

The restrictions on the selection of the objects that we could observe in the allocated time meant that classical FR~II galaxies were ultimately excluded because their cores are typically weak. One exception is PKS~0859--25 due to the fact that, unfortunately, the data were   overly affected by radio frequency interference (RFI). Thus, this selection has introduced a bias in the observed sample which will be considered in the discussion of the results.
The types of sources in the present sample are very different from those studied in \cite{Morganti01} for the low-$z$ ($z<0.2)$ part of the 2-Jy sample. That sample included only four broad-line objects (and one BL Lac) and instead had a number of weak emission-line galaxies. This already illustrates some of the differences that are present when moving to relatively high-$z$ radio sources.
The targets of the present study are also strong radio sources as shown in the plot of the radio luminosity versus  redshift of Fig.\ \ref{fig:Distribution-z}, modified from \cite{Murthy22},  where the comparison of  the distribution of radio luminosities with those of other studies can be found. 

For the discussion of the sample, it can also be useful to have the UV luminosities of the sources, considering the role expected it may play according to some studies (see Sect. 1). \cite{Curran08} derived the UV luminosities at 1216\,\AA\, by extrapolation or interpolation of mostly the optical $B$ and $R$  band fluxes and subsequent studies have followed this approach. In the case of our sample, some objects have been observed by GALEX, while for the rest we extrapolated to 1216\,\AA\, from the shortest wavelength UV/optical wavelength for 
which we could find a flux measurement in the literature. We list the UV luminosities in Table \ref{tab:parameters} and we describe how they were derived in Appendix \ref{sec:AppendixUV}.

\begin{figure*}
\centering
\includegraphics[width=0.95\linewidth]{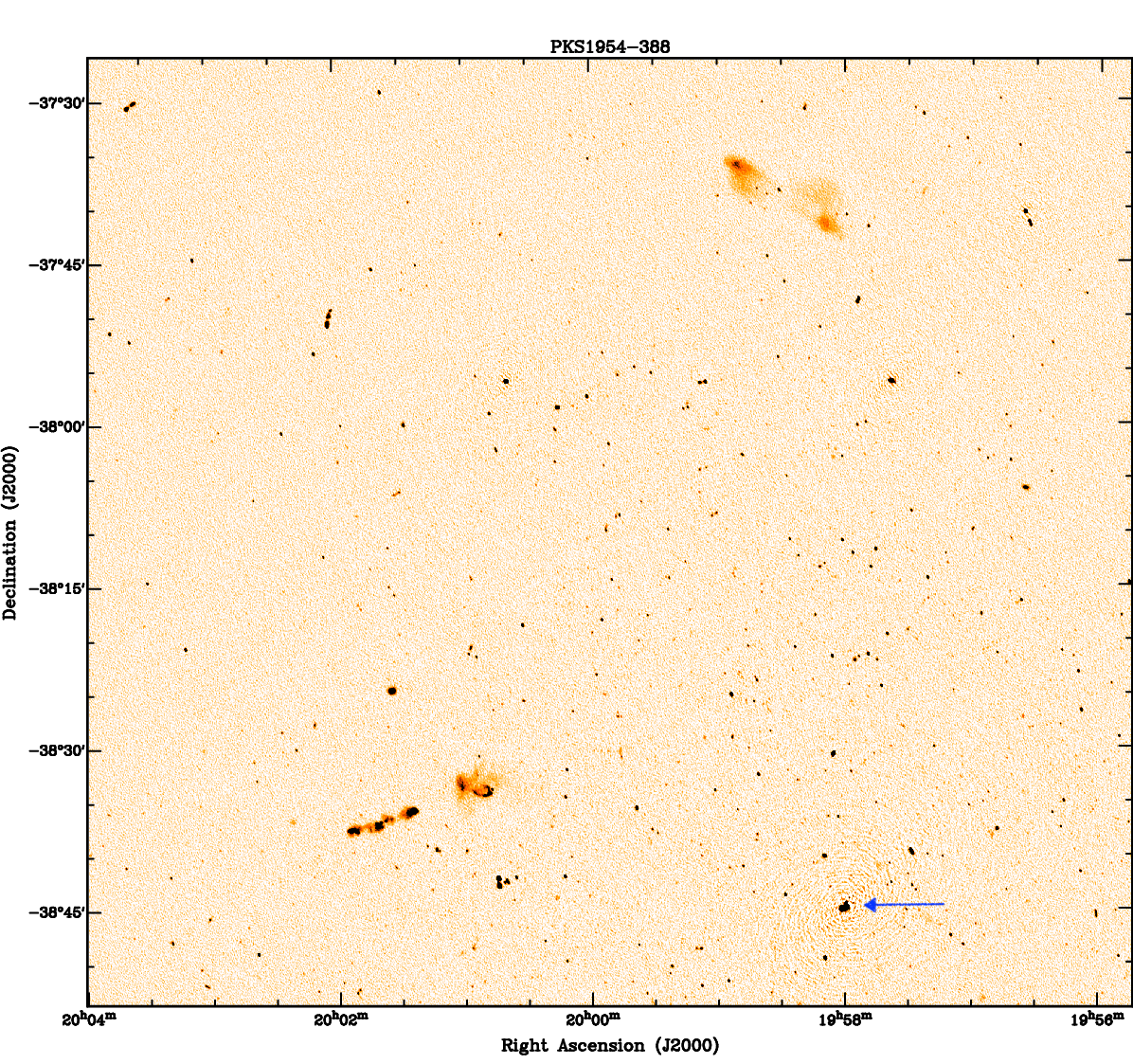}
\caption{\label{fig:PKS1954Bello} Example of one of the continuum images obtained (PKS~1954--38), illustrating the image quality. The target, PKS~1954--38, is visible in the bottom of the image indicated with an arrow,  and the residual artifacts around this source are due to the strength of the source. The total flux of PKS~1954--38 is 1.195 Jy and its peak is 940 \mJybeam. The root-mean-square level of the artefacts around PKS~1954--38 is 0.1 \mJybeam\ while the noise level away from the target is 0.05 \mJybeam. This indicates that the dynamic range near the target is about $10^4$.}
\end{figure*}

\section{MeerKAT observations and data reduction}
\label{sec:Observations}

The observations were carried out with MeerKAT in October and November 2022 under  project 20220822. Although  the redshifts of the sources were such that  their \HI\ line would not fall  in   a priori known RFI regions, the data of four objects were affected by RFI. In two cases (PKS~0842--75 and PKS~0859--25), the spectra turned out to be not usable, while in one case (PKS~1136--13), the RFI was intermittent and the data could be used after flagging. In one final case (PKS~1151--34), the RFI  only affects part of the spectrum and avoids the range of redshifted \HI. 
Thus, the final usable sample comprises 15 sources.

The sample sources were observed in five blocks, four in the UHF band and one with the L-band receiver, depending on the redshift of the sources. The observations were done with 62 antennas, and each target was observed for 1 h on-source. The UHF data cover the range 544--1088 MHz, while the L band covers the range 856--1712 MHz.  The 32K mode of the SKA Reconfigurable Application Board (SKARAB) correlator has been used. We used either PKS 1934--63 or J0408--6545 as primary calibrator for the flux density scale, delay, and bandpass calibration. Given the fact that our targets are quite strong sources, each observation of a primary calibrator was 30 minutes in length; in addition, in each observation block of 5 h, a bandpass calibrator was observed at the start and at the end of the block in order to monitor variations in the bandpass calibration (see Appendix \ref{sec:AppendixBP} for an example).
The frequency of the redshifted \HI\ for each source is listed in Table \ref{tab:obs}, together with the observational details and the parameters of the final data products.

In each observing block of 5 h, several sources were observed using the full UHF- or L bandwidth. However, to reduce the data volume for the analysis, for each source  a band of about 20 MHz was extracted, centred on the frequency of the redshifted \HI. This small band was also used for obtaining the continuum images (see below). The full band can be used to do a blind exploration of the presence of \HI\ in other radio sources in the field, which we will present in a future paper.

All calibration steps were done using the CASA software \citep{McMullin07}, except for cleaning the images, which was done using WSClean \citep{Offringa14}, and for subtracting the clean components from the visibilities, which was done using DP3 \citep{Dijkema23}. The first step in the calibration was to determine the delay calibration using one of the primary calibrators. Next, a initial bandpass and flux calibration was done for the primary calibrators using a point source model. Using this initial calibration, images were made of the primary calibrators, which were self-calibrated in a standard way. The source model resulting from  this was used as  model in a second iteration for deriving an new flux and bandpass calibration, now taking other sources in the field into account as well. This second step gives an improvement of the  bandpass calibration, which is needed because of the strong fluxes of our target sources.

For each target source, the calibration of the primary calibrator observed closest in time was used to cross calibrate the target observation. Given the high fluxes of the targets, no secondary calibrators were used in the calibration. Each target observation was
self-calibrated iteratively using three iterations of phase calibration, followed by one iteration of amplitude calibration. Only direction-independent calibration was performed. The continuum images obtained in this way (using robust = -- 1.5) have a typical resolution of 10 arcsec, a noise level varying between 0.07 and 0.6 \mJybeam\ and a typical dynamic range of  a few times $10^4$. The exception is PKS~0842--75, for which (due to RFI) a continuum image could not be generated (see Table \ref{tab:obs}). Figure\ \ref{fig:PKS1954Bello} shows an example of a continuum image obtained in this study, where the high quality is clearly seen. 

Given the large field of view of MeerKAT, in particular in the UHF band, the continuum subtraction has to be done by subtracting the sky model from the target visibilities. After this was done, a data cube was made using robust = 0 and with a velocity resolution of about 10 \kms (varying with redshift; see Table \ref{tab:obs})  covering a small area centred on the target source. Although the calibration and the sky model obtained are of high quality,  some minor continuum residuals are still present in these data cubes, typically at the level of a few  times the noise level. These residuals were subtracted from the data cubes using a Savitzky–Golay filter on the spectra. The noise level of the final data cubes is typically 1 \mJybeam.

As mentioned above, for one source (PKS~0842--75) the continuum image could not be made due to strong RFI affecting the band. 
Ten sources show extended radio continuum (listed in column 12 of table \ref{tab:parameters}), with cases of double-lobed structure and other sources where only a small extension is seen. In the vast majority of the sources, a prominent core is observed (as expected). The continuum images of the extended sources are shown in Appendix \ref{sec:AppendixFigRadioCont}.

\begin{table*}
\caption{Parameters of the sources observed with MeerKAT.  }
\tiny
\begin{center}
{\setlength{\tabcolsep}{5pt}
\begin{tabular}{lcccccccccccc}\\
\hline\hline
    Source            &  $S_{\rm tot}$ & $\log L_{\rm tot}$ & $S_{\rm core}$ & $\Delta S_{\rm peak}$ & $\tau_{\rm peak}$ & $\int S\,dv$ & $W_{50}$ & $W_{20}$ & $W_{0}$ & $N_{\rm \HI}$  & Morph & $L_{\rm UV}$ 1216\AA\\
                &  (Jy)     & (\WHz)     & (Jy)      &    (mJy)       &    ($10^{-3}$)  & (mJy \kms) &            (\kms) & (\kms) & (\kms) &  ($10^{20}$ cm$^{-2}$) & &  ($10^{23}$ \WHz) \\
\hline
0159--11    &  3.615 &   28.05  & 3.280     &         &$<0.59$ &    & & &   & $<0.07$ & E & 120\\
0252--71    &  8.879 &   28.24  & 8.879     &         &$<0.59$ &    & & &   &  $<0.07$   & U & 0.02\\
0403--13    &  4.964 &   28.00 &  4.964      &         &$<0.51$ &    & & &   &  $<0.09$   & U & 1.9\\
0405--12    &  4.132 &  27.93 &  0.839\rlap{$^a$}  &  \phantom{0}$4.5\pm1.0$\rlap{$^b$}  &  ~~~~4.76  & \phantom{0}$372\pm 29\phantom{0}$ & $90\pm 20$ & $125\pm 30$ & $160\pm 20$ & ~~~0.81     & E & 22\\
0637--75    &  6.938 &  28.30 &  4.980      &         &$<0.64$ &    & & &   &   $<0.12$  & E & 9.5\\
0842--75    &   --   &   --   & --    &         & --    &      & & &   &   --    & E  & 17\\
0859--25    &  7.539 &  27.45  & --          &         & --    &       & & &   &  --    & E  & 0.07\\
1136--13    &  6.332 &  28.00 & 2.280       &         &$<1.20$ &     & & &   &  $<0.22$  & E & 3.8\\
1151--34    &  7.498 &  27.27 & 7.498       &   $28.0\pm2.0$    &  ~~~~4.0   & $5386\pm 105$ & $157\pm 20$ & $350\pm 25$ & $450\pm 30$ &   ~~~1.31     & U & 0.18\\
1306--09    &  5.429 &  27.79 & 5.429      &  \phantom{0}$4.5\pm1.0$      &  ~~~~0.9   &  $1072\pm 69$\phantom{0} &  $260\pm 30$ &  $405\pm 30$ & $455\pm 30$ &  ~~~0.36    & U & 0.08\\
1355--41    &  5.905 &  27.38 & 0.115      &          & $<17.1$ &   & & &   &  $< 3.11$  & E & 4.2\\
1510--08    &  3.332 &  27.29 & 1.470      &         & $<1.41$  &   & & &   &  $<0.26$   & E & 0.60\\
1602+01     &  6.418 &  27.86 & 3.220       &         & $<0.90$ &   & & &   &  $<0.16$   & E & 0.81\\
1954--388   &  1.195 &  27.50 & 0.947       &         & $<1.74$ &   & & &   &  $<0.32$   & E & 1.20\\
2203--18    &  7.288 &  28.26 & 7.288      &         & $<0.49$ &    & & &   &  $<0.09$   & U & 0.04\\
2243--123   &  1.692 &  27.65 & 1.580      &         & $<1.12$ &    & & &   &   $<0.20$  & E & 2.10\\
2345--16    &  2.602 &  27.73 & 2.602      &         & $<0.89$ &    & & &   &   $<0.16$  & U & 0.71\\
\hline
\end{tabular}
}
\tablefoot{The continuum fluxes are obtained at the frequency of the \HI\ observations; therefore, at 1.4 GHz in the rest frame at the redshift of the source and it is the same for the total luminosities. The column densities are assuming \tspin = 100 K and a covering factor $c_{\rm f}$ = 1. The upper limits to the column density are estimated using the 3-$\sigma$ noise and a FWHM of the line of 100 \kms. $^{(a)}$ Flux of the peak of the southern lobe. $^{(b)}$ Absorption against the southern lobe.  }
\end{center}
\label{tab:parameters}
\end{table*}

\section{Results}
\label{sec:Results}

\begin{figure*}
\sidecaption
\includegraphics[width=0.33\linewidth]{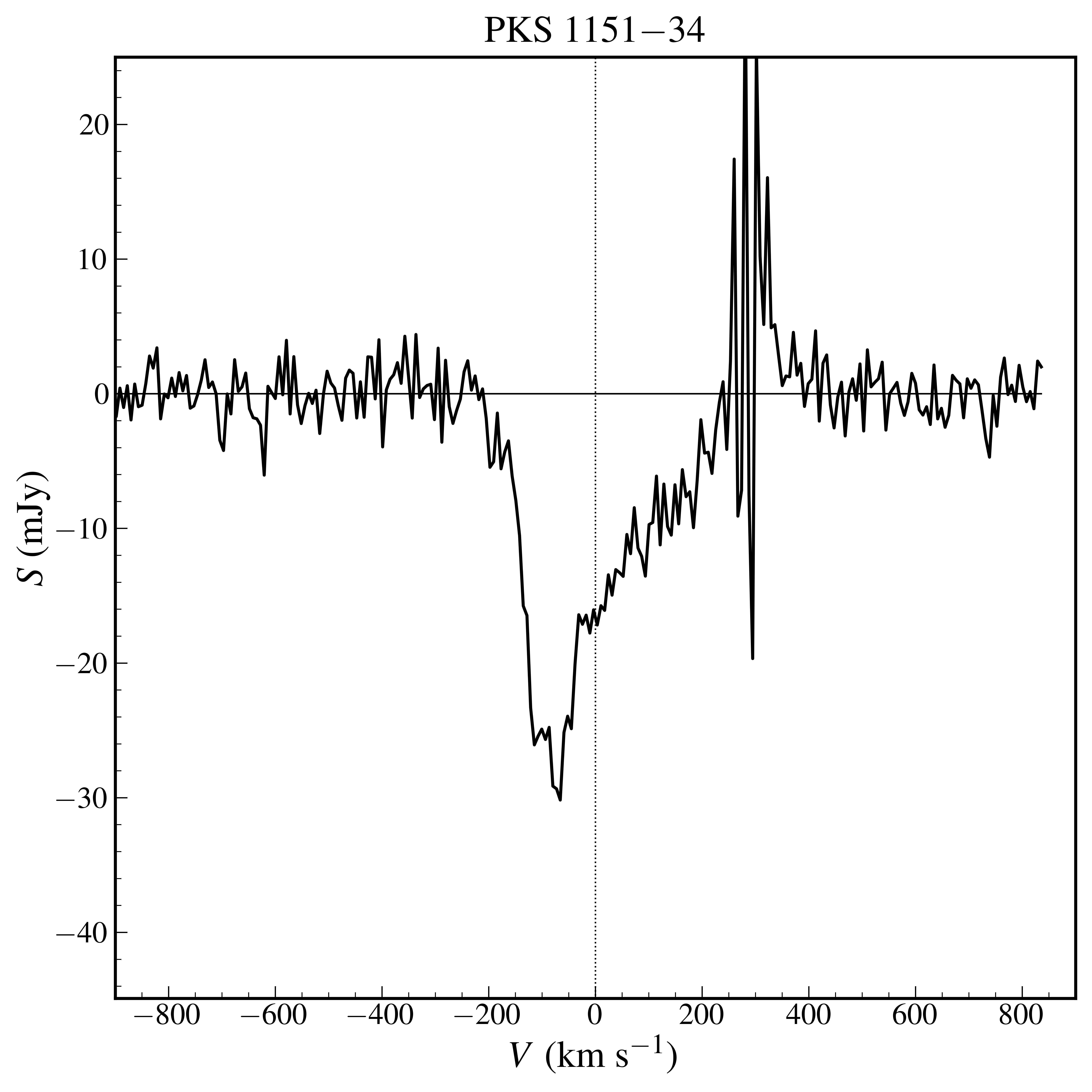}
\includegraphics[height=0.33\linewidth]{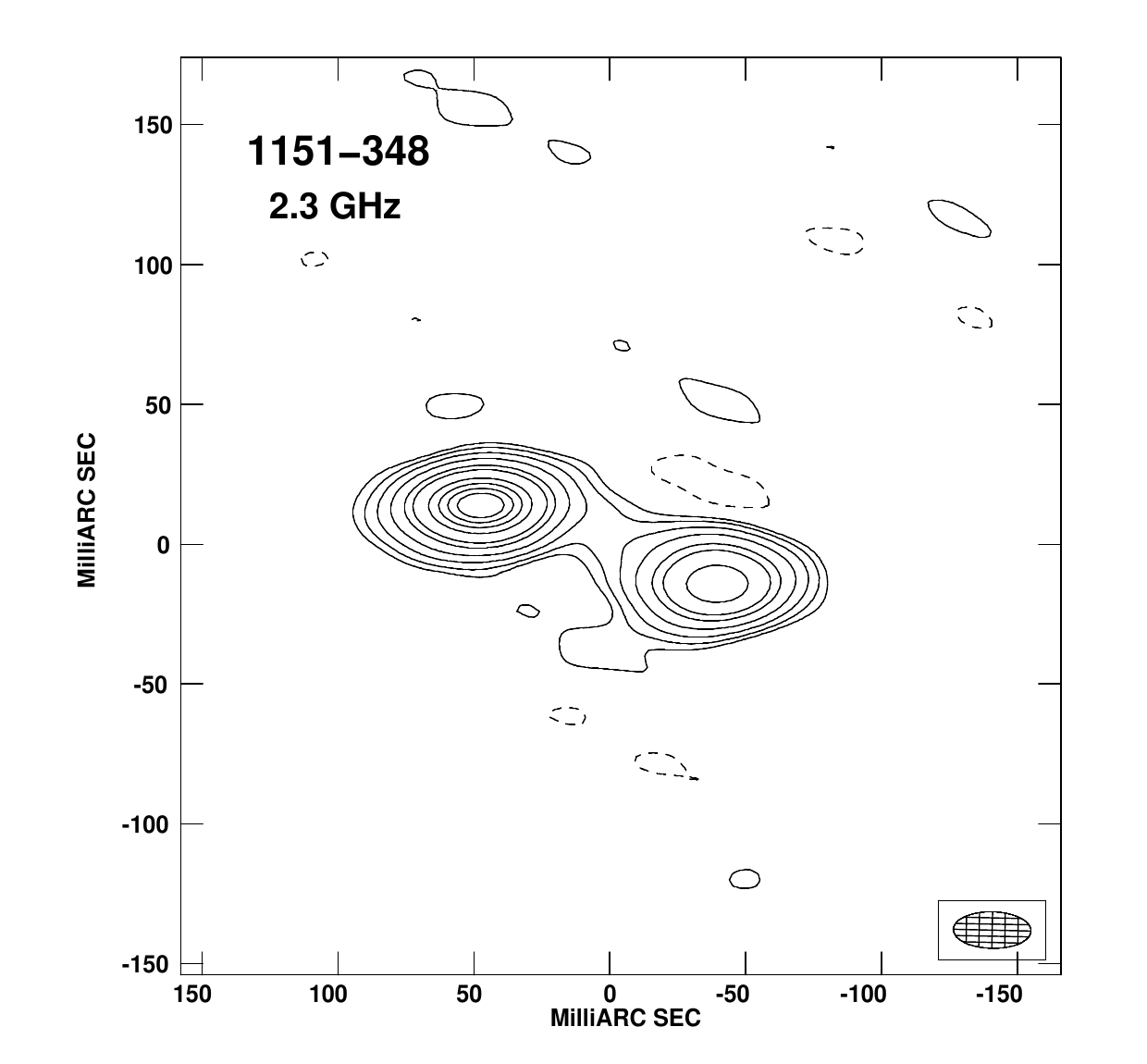}
\caption{  {Left panel:}  \HI\ absorption profile in PKS~1151--34. Note: the RFI affecting part of the band. The error in the zero-point of the velocity scale is 30 \kms\ (see Table \ref{tab:obs}). Right panel: VLBI image of PKS~1151--34  from \cite{Tzioumis02}.  The continuum emission spans $\sim$700 pc.
}
\label{fig:TwoDetections1}
\end{figure*}

\begin{figure*}
\sidecaption
\includegraphics[width=0.33\linewidth]{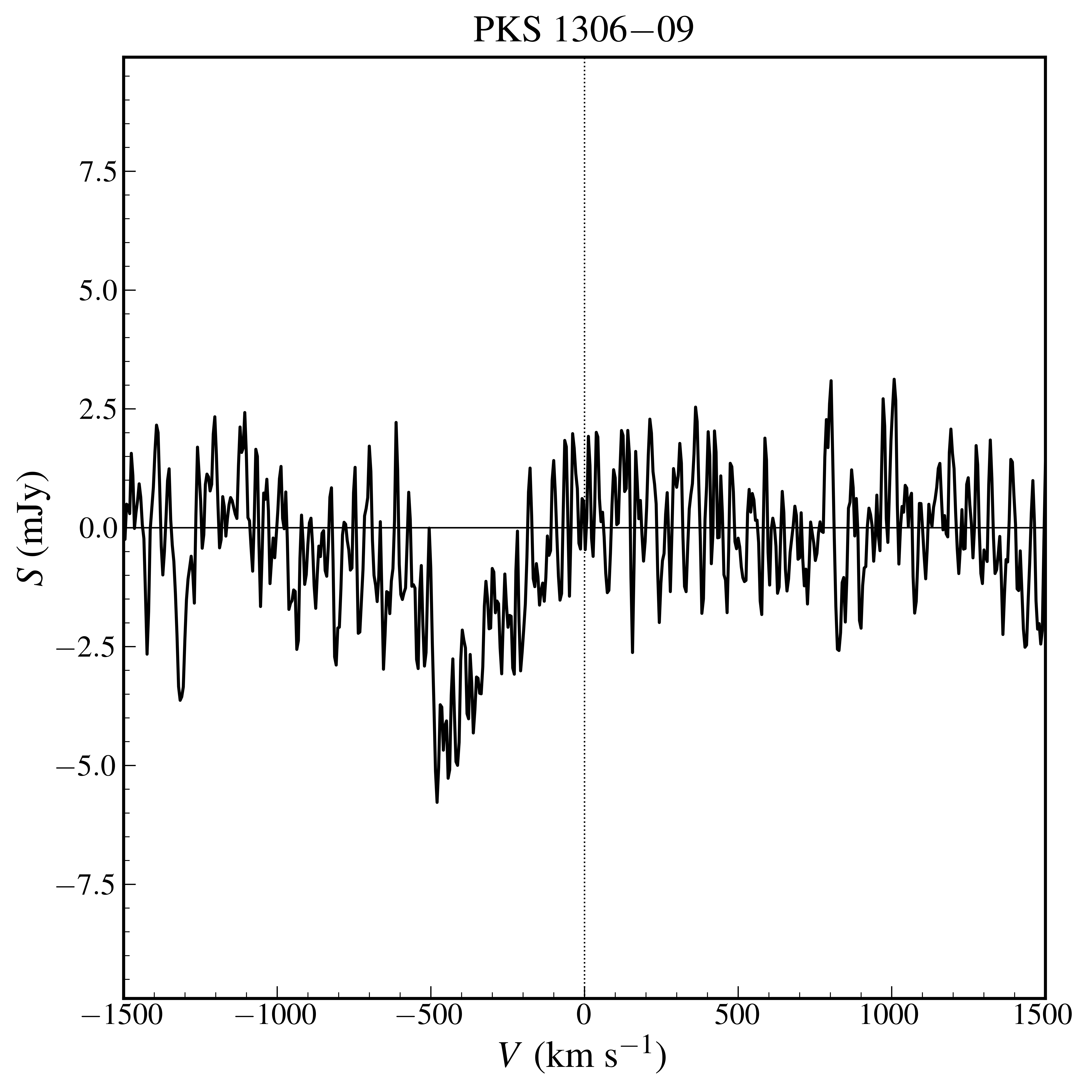}
\includegraphics[height=0.33\linewidth]{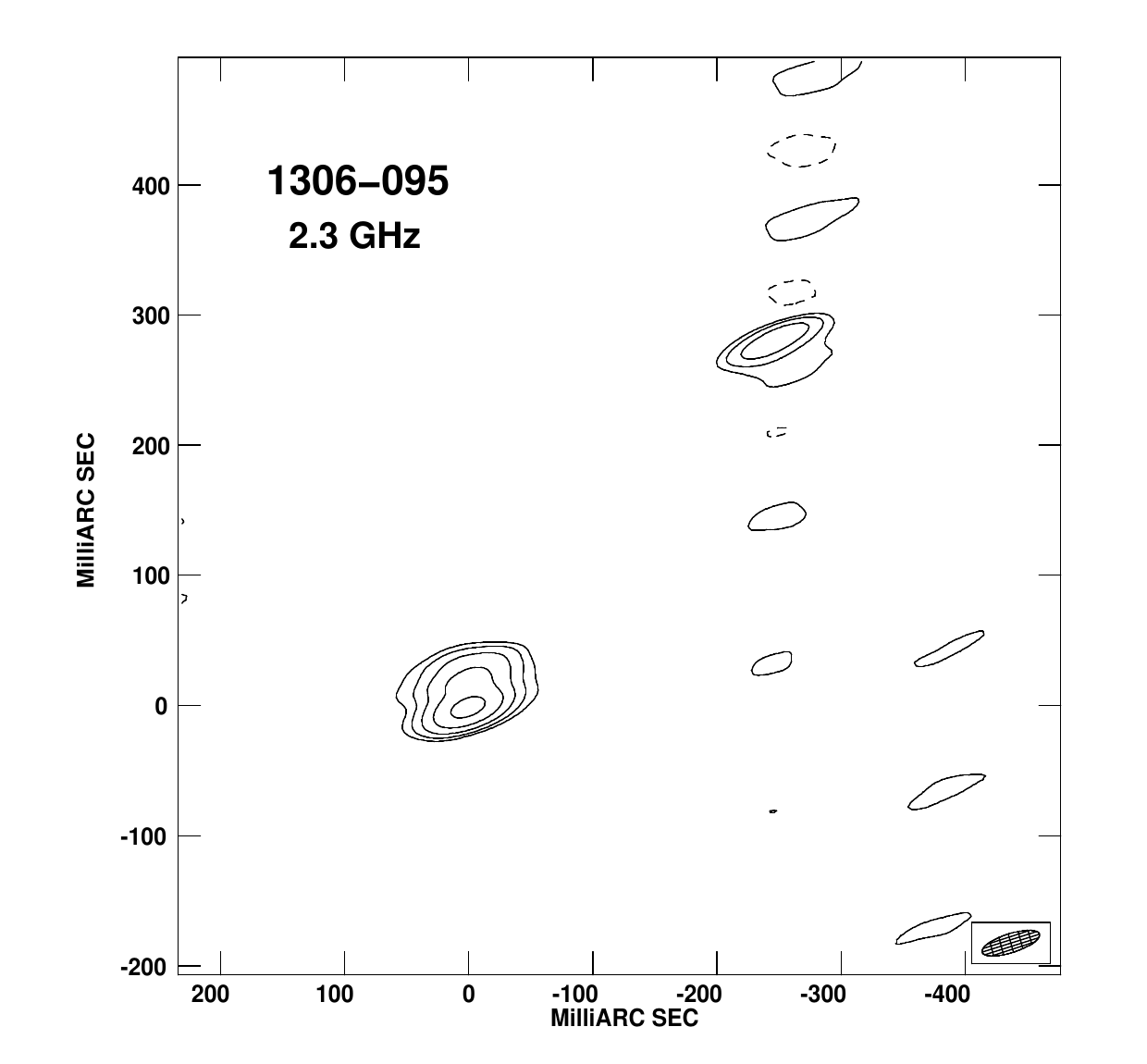}
\caption{  {Left panel:} \HI\ absorption profile against the young radio galaxy PKS~1306--09. The error in the zero-point of the velocity scale is 20 \kms\ (see Table \ref{tab:obs}). Right panel:   VLBI image of PKS~1306--09 taken from \cite{Tzioumis02}. The size of the continuum emission is 2.4 kpc. 
}
\label{fig:TwoDetections2}
\end{figure*}

\begin{figure*}
\sidecaption
\includegraphics[width=0.33\linewidth]{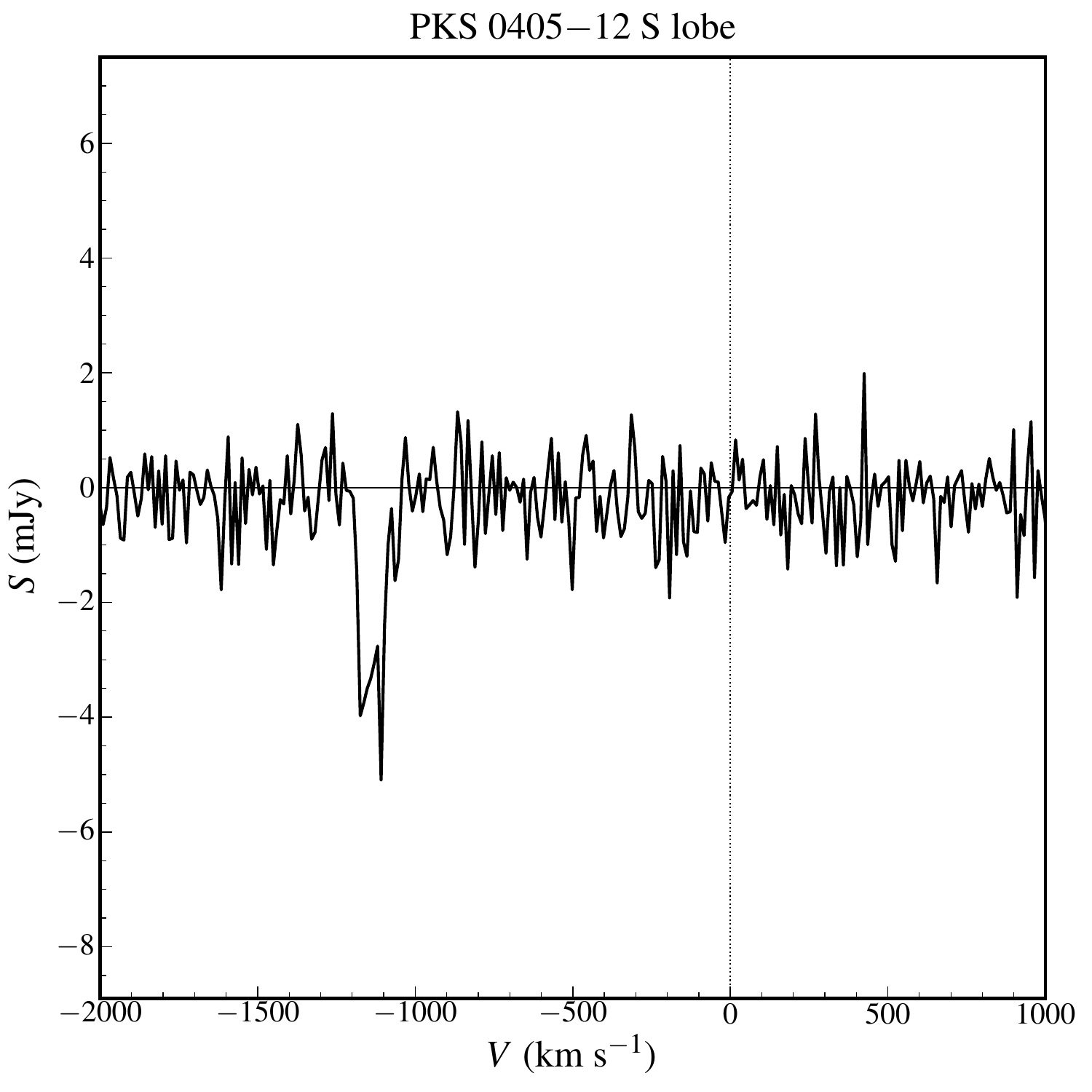}
\includegraphics[width=0.345\linewidth]{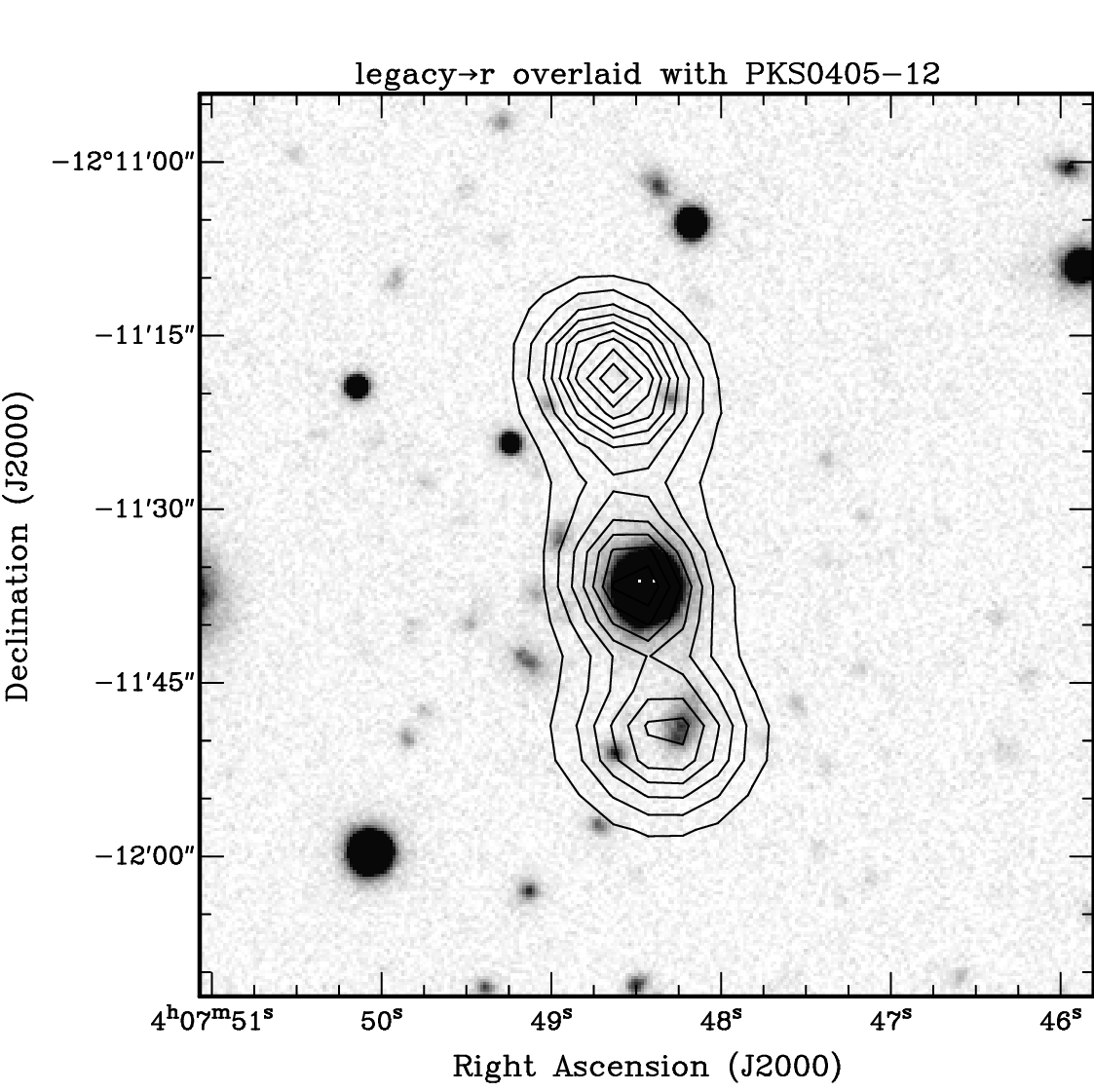}
\caption{ {Left panel:}  \HI\ absorption detected against the southern lobe of PKS~0405--12 where an optical galaxy is also visible.  {Right panel:} Continuum image of PKS~0405--12 superposed to the optical r-band image of PKS~0405--12 from DESI Legacy Imaging Surveys \citep{Dey19}.  }
\label{fig:0405}
\end{figure*}

The results from our search for \HI\ absorption are summarised in Table \ref{tab:parameters}.
Of the 15 sources for which the effects of RFI were negligible or limited, we detected \HI\ absorption in three objects. Two are associated \HI\ detections (PKS~1151--34 and PKS~1306--09). These two detections  result in a detection rate of 13\% $\pm$ 7\% for the associated absorption.  We defined the third detection (PKS~0405--12)   as 'local intervening' because it is attributed to a foreground galaxy in the local environment of PKS~0405--12. The de-redshifted \HI\ spectra of the detections are shown in Figs \ref{fig:TwoDetections1}, \ref{fig:TwoDetections2}, and \ref{fig:0405}.

\begin{figure}
\centering
\includegraphics[width=0.95\linewidth]{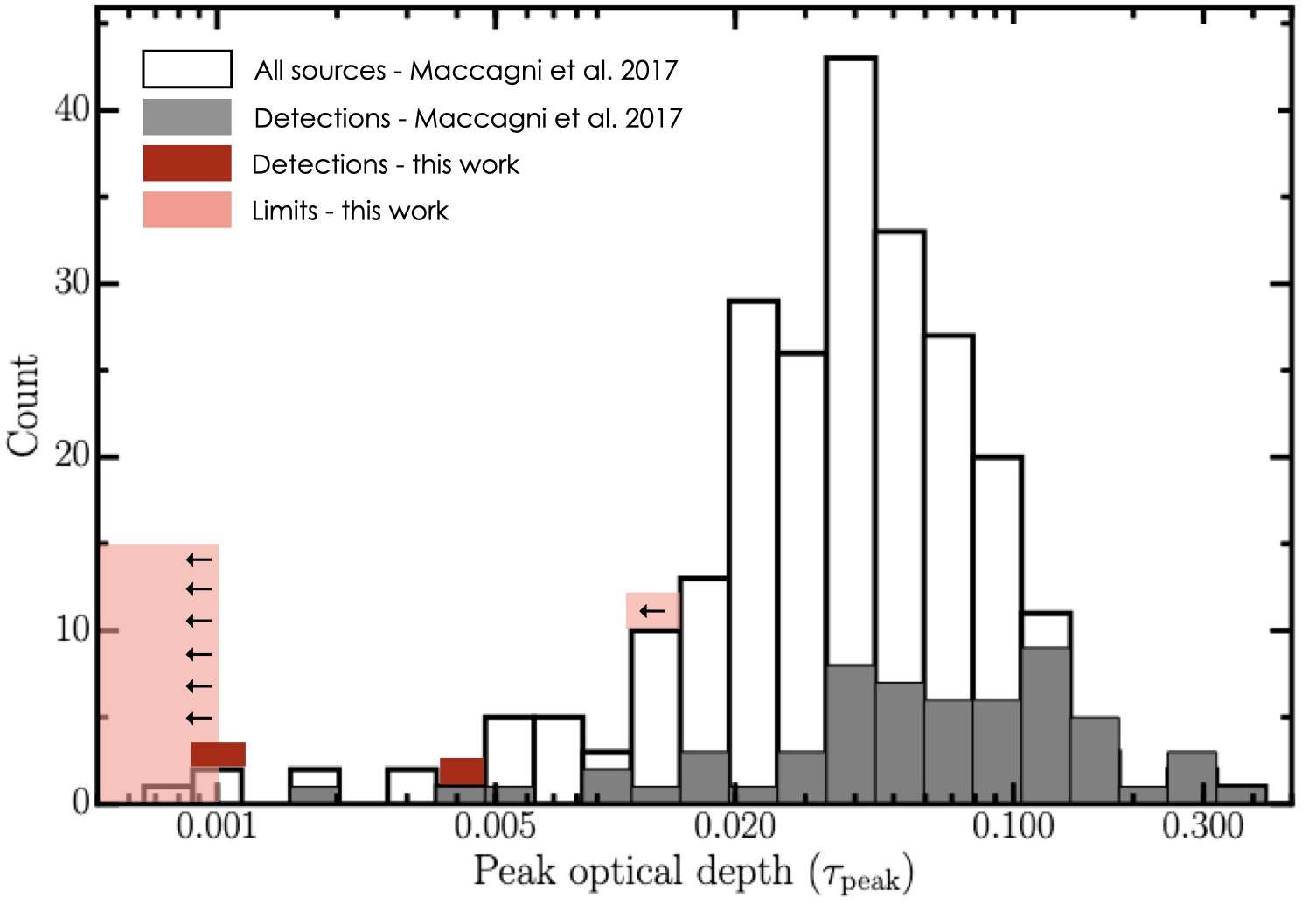}
\caption{\label{fig:Distribution-tau} Distribution of the peak optical depth taken from \citet[grey and empty bars]{Maccagni17} with overlaid the peak optical depth for the detected (dark red boxes) and limits thereof (light red) for the sources of the present sample. The distribution shows how the present observations reach very low optical depths.}
\end{figure}

Interestingly, the two associated \HI\ absorptions were found against CSS and peaked (i.e.\ young) radio sources, despite the fact that one of them (PKS~1151--34) is classified as quasar (see Sect.\ \ref{sec:1151}). 
These detections confirm the trend, which was already found at both low and high redshift, that young radio sources are more likely to be detected in \HI\ (see also Sect.\ \ref{sec:Discussion}). The other two CSS and peaked  sources in the sample were not detected. 

The peak optical depth (given in column 6 of Table \ref{tab:parameters}) is defined as $\tau_{\rm peak} = -\ln[1-\Delta S/(S_{\rm cont} \times c_{\rm f})]$, where $\Delta S$ is the absorbed flux, $S_{\rm cont}$ the continuum flux (either total or core depending whether the source is unresolved or extended), and $c_{\rm f}$ is the covering factor, assumed to be unity. In  cases of non-detection, the upper limit to the peak optical depth has been calculated using  $3 \times \sigma_{\rm line}$. In the case of the local intervening detection in PKS~0405--12, the continuum flux used to derive $\tau_{\rm peak}$ is the one measured at the location of the absorption.

Looking at the values listed in Table \ref{tab:parameters}, we can see, due to the combination of the strength of the radio continuum of the sources (and their cores) and the high sensitivity of the observations, the  limits of  $\tau_{\rm peak}$  are quite low, in many cases,  $\tau_{\rm peak}$ is even smaller that 0.001 (0.1\%). Also the two detections of associated absorption have low optical optical depths between 0.004 and 0.0009 (0.4\% and 0.09\%).
This is illustrated in Fig.\ \ref{fig:Distribution-tau} where the $\tau_{\rm peak}$ of the objects in our sample are compared with literature findings taken from \cite{Maccagni17}, which is representative of the variety among low-$z$ radio sources. Although some cases of such low optical depth are also seen at low-$z$, the detections in our sample have $\tau_{\rm peak}$ values that are well below the peak of the distribution of the low-$z$ detections. In \cite{Maccagni17}, only 3 out of 248 objects have $\tau_{\rm peak} \lesssim 0.005$. Interestingly, two of these are young radio sources. Also our upper limits cover a very different range compared than the distributions shown in Fig.\ \ref{fig:Distribution-tau}.

For the detections, Table \ref{tab:parameters} lists, in addition to $\tau_{\rm peak}$ values, the width of the absorption lines and the column densities.
The values of the column densities presented in Table \ref{tab:parameters} are normalised for \tspin\ and covering factor. 
The integrated column density has been estimated as $N_{\rm \HI} = 1.83 \times 10^{18} T_{\rm spin}/c_{\rm f} \int \tau dv$. For the non-detections, a width of 100 \kms\ has been assumed.
For the usually assumed values of \tspin = 100 K and $c_{\rm f}=1$, the upper limits to the column densities are relatively low, below $10^{20}$ cm$^{-2}$. We discuss these assumptions, in particular \tspin\ for PKS~1151--34, below.

\subsection{Young radio sources detected in H\,{\small I}}
\label{sec:DetectionsYoung}

The associated \HI\ detections are found in two of the  young (CSS) radio sources: PKS~1151--34 and PKS~1306--09. A discussion of their radio properties is provided in \citep{Tzioumis02}. Their \HI\ profiles are shown in Figs \ref{fig:TwoDetections1} and \ref{fig:TwoDetections2}. 
In both sources, the warm ionised gas has been studied in detail,  recently using  X-shooter by \cite{Santoro20}.  As often seen in young radio sources, both show kinematically disturbed ionised gas associated with outflows. The most extreme and extended outflow is observed in PKS~1306--09. In this source, the fast outflow is aligned with the direction of the radio plasma and has comparable extent. The emission lines are broad (the FWHM of the \OIII$\lambda$5007 line is $\sim$1000 \kms) and double-peaked, with the blueshifted component of \OIII$\lambda$5007 having a velocity of $\sim$500 \kms.   The situation in PKS~1151–-34 is more difficult to characterise because of the presence of broad permitted emission lines in the spectrum. However, in both cases the (outflowing) ionised gas appears to be characterised by high densities (log $n_e \sim 3 - 4$ cm$^{-3}$). All this suggests the presence of an interaction between jets and gas -- often seen in young radio sources -- which could result in compression
of the gas by jet-induced shocks.

The \HI\ absorption profiles of the two sources show different characteristics which are discussed below.  
Both sources are particularly strong in radio continuum (about 7 and 5 Jy, respectively), while the \HI\ absorbed flux is faint. The detection of such faint absorptions (especially in PKS~1306--09) confirms the good quality and stability of the bandpass of the MeerKAT telescope (more details are given in Appendix \ref{sec:AppendixBP}).

\subsubsection{A young quasar: PKS~1151--34}
\label{sec:1151}

In PKS~1151--34, we detected \HI\ absorption with a broad profile  ($W_{50} = 157$ \kms and $W_{20} = 350$ \kms) and peak optical depth of only $\tau_{\rm peak} = 0.004$. The absorption profile is roughly centred on the systemic velocity, but it is asymmetric in intensity, with the deeper peak on the blueshifted side  (blueshifted by $79 \pm 2$ \kms). The asymmetry does not appear to be due to the presence of RFI on the redshifted side of the spectrum (although the exact width on the redshifted size could be affected by the RFI). However, it is more likely that it represents the shape resulting from the distribution of the gas combined with the morphology of the background continuum (see below). 

As mentioned above, PKS~1151--34 is a young radio source. This is indicated by the  steep spectrum,  combined with the radio morphology and size. The radio spectrum is steep at high frequencies (with spectral index $\alpha^{\rm 2.7~GHz}_{\rm 8~GHz}=-0.7$, defined through $S\propto \nu^{\alpha}$, with $S$ the continuum flux; see \citealt{Tzioumis02}) with a flattening at lower frequencies, possibly indicating the separation between optically thin and thick emission.  The source has a double-lobed structure with a total projected extent of about 680 pc. 
The VLBI image taken  from \cite{Tzioumis02} is shown in  Fig.\ \ref{fig:TwoDetections1} (top-right). Neither of the two VLBI components shows the flat type of spectrum that would otherwise be typical of a core.  These properties support the scenario in which the source is small because it is young (and not because of any extreme orientation \opm{or beaming effects, also} considering the relatively symmetric lobes), despite the optical classification.

\begin{figure}
\centering
\includegraphics[width=0.8\linewidth]{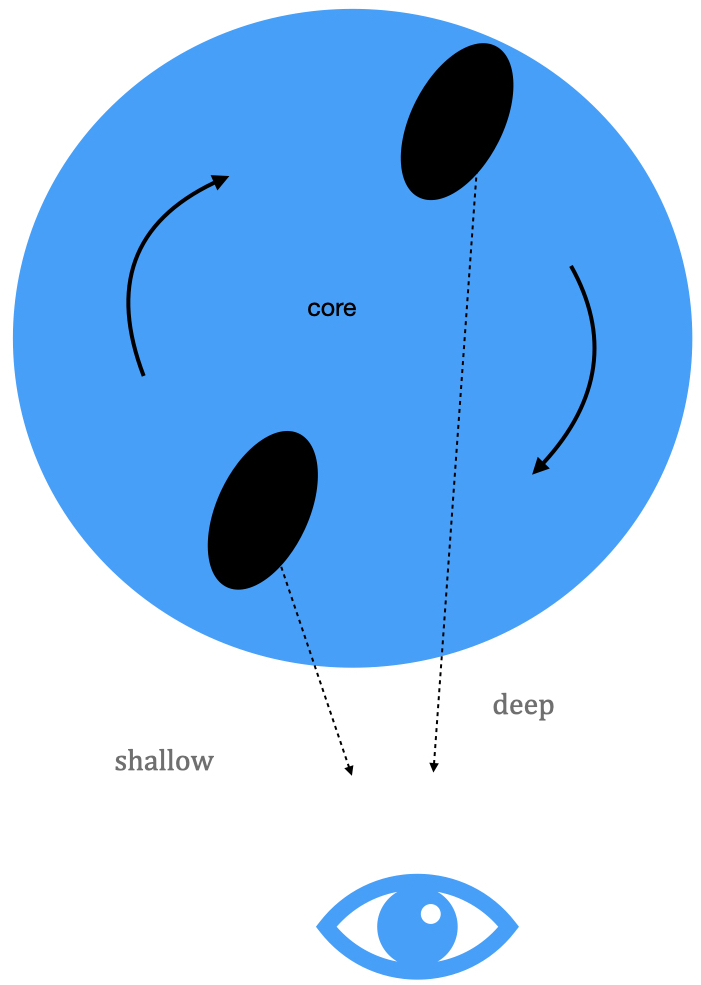}
\caption{ Cartoon of the possible geometry explaining the absorption in the young quasar PKS~1151--34 (Sect. \ref{sec:1151}). 
}
\label{fig:1151-Model}
\end{figure}

At optical wavelengths, PKS~1151--34 has been classified as a broad-line radio galaxy based on the detection of broad Balmer lines of moderate strength, although its \OIII$\lambda$5007 emission-line luminosity is more consistent with that of a quasar \citep{Santoro20}. The clear presence of stellar absorption
lines in its nuclear spectrum and the relatively red colours suggest that its AGN nucleus may be partially obscured at optical wavelengths.

The \HI\ absorption could originate from the radio source being embedded in a circumnuclear gas disc.  This is suggested by the fact that the absorption is roughly centred on the systemic velocity and by the shape of the absorption with a sharp edge of the profile on the blueshifted side, similar to those seen in emission for rotating discs and rings. The asymmetry in intensity could be due to the orientation of the radio source relative to the disc and ring, as schematically shown in the cartoon in Fig.\ \ref{fig:1151-Model}. 
In this scenario, the deeper, blueshifted absorption would originate from the lobe located in the background (i.e.\ crossing a larger layer of gas). 
If the \HI\ profile is due to gas rotating in a disc, the observed width of the profile ($W_0=450$ \kms) at a radius of 340 pc (i.e. the location of the radio lobes) can be explained by a central mass of the order of $4 \times 10^9$ \msun\ (without considering the effect of inclination). The mass of the BH in the centre of PKS~1151--34 can be estimated from the stellar mass-black hole mass relation (e.g.\ \citealt{Kormendy13}).
The upper limit to the stellar mass (uncorrected for AGN contamination of the K-band luminosity) is $4\times10^{11}$ \msun,  as derived by  \cite{Bernhard22}. This results in an upper limit to the BH mass of $2.5\times 10^9$ \msun. Thus, given the uncertainties, the rotation of the gas can be sustained by the central mass. 
The full confirmation of this scenario would require the location of the \HI\ absorption in the radio lobes to be spatially resolved. Furthermore, the presence of gas with kinematics deviating from regular rotation, such as in- or outflowing clumps of gas, cannot be excluded. For this, observations with the southern VLBI array (Australian
Long Baseline Array, LBA) reaching the frequency of \HI\ at the redshift of PKS~1151--34 (i.e.\ 1129.09 MHz) will be needed. 

\begin{figure}
\centering
\includegraphics[width=0.9\linewidth]{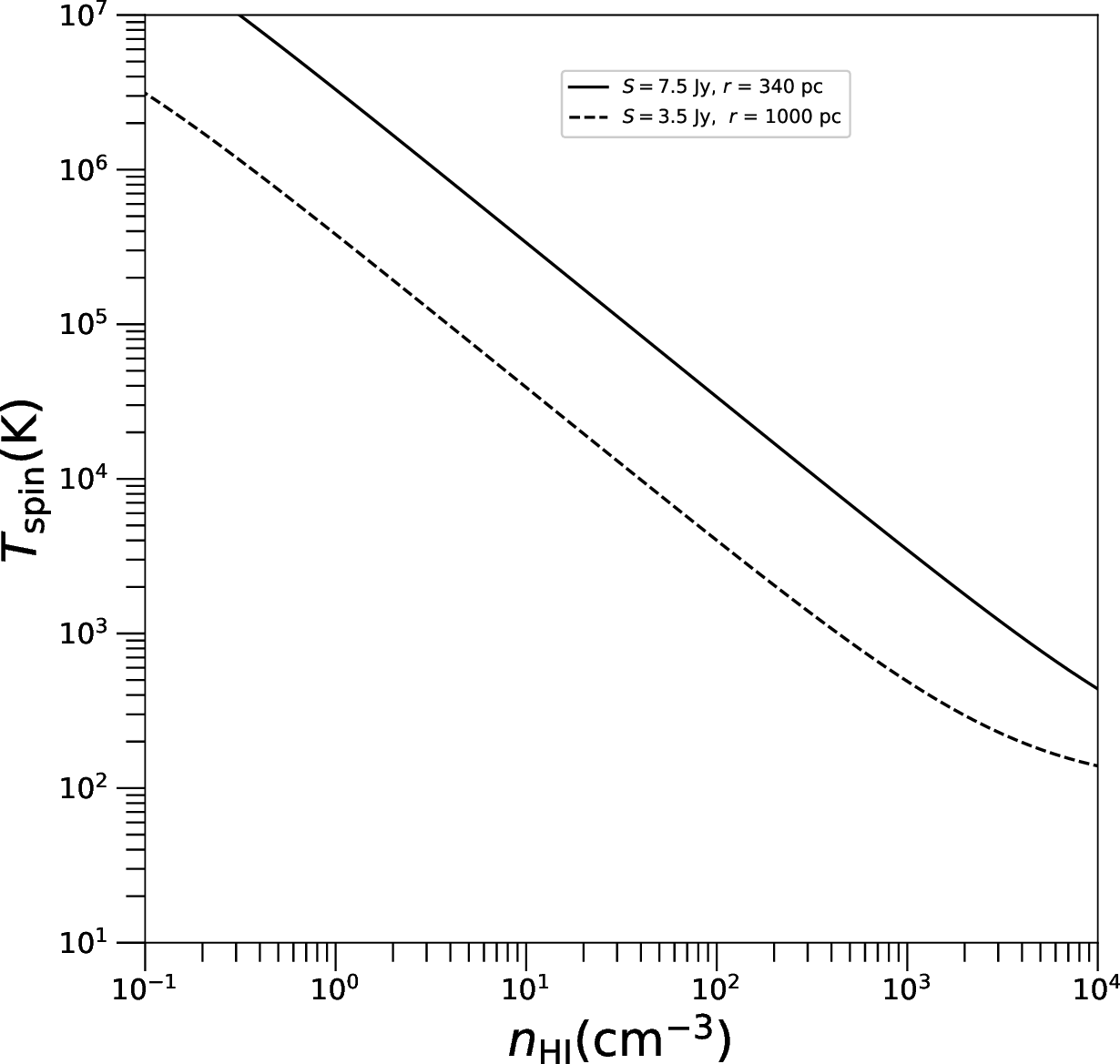}
\caption{ Estimated \tspin\ as function of the column density derived for PKS~1151--34  following the calculation in \cite{Bahcall69}; see Sect.\ \ref{sec:1151} for details. The plot shows that high values of \tspin\ are expected for the typical densities of the \HI\ clouds.
}
\label{fig:1151-tspin}
\end{figure}

Pinpointing the location of the absorption will also clarify whether all the \HI\ is part of a regularly rotating structure or whether turbulent gas motions are also present (as hinted at by the broad absorption profile), possibly due to the jet interacting  with the gas, as in the case of 4C~31.04 \citep{Murthy24}, and/or resulting from gas settling following accretion from a companion galaxy. This source has not been observed by GALEX (it is outside the area covered by GALEX). However, thanks to the X-shooter observations  of \cite{Santoro20}, the UV luminosity could be estimated (see Appendix \ref{sec:AppendixUV}). This was done by extrapolating the measured  fluxes  from the Xshooter spectra at rest-frame wavelengths $\sim$2200 -- 2700 \AA\ to 1216 \AA\ using the \cite{Berk01} mean quasar SED. Such an extrapolation is uncertain, given the (likely significant) contribution of starlight in the near UV. Therefore, the derived luminosity should be considered an upper limit. The UV luminosity derived in this way is $L_{\rm UV} = 1.8\times 10^{22}$ \WHz, which is below the limit proposed by \cite{Curran08}; therefore, the UV radiation is not expected to be hostile to the presence of \HI. However, PKS~1151--34 is a strong radio source and, as described in the introduction, 
this can also have an effect on the conditions of the gas and, in particular, on the \tspin, which, in turn,   affects the observed optical depth. To illustrate the possible impact of this, we can use the formulae presented in \cite{Bahcall69} to estimate the \tspin\footnote{Note that a theoretical upper limit of 8000 K has been estimated by \citealt{Maloney96} above which all \HI\ is expected to be ionised}. 
The results are shown in Fig.\ \ref{fig:1151-tspin} where we plot \tspin\ as function of \HI\ volume density for two scenarios. 
Based on the ides of the radio lobes being embedded in the gas disc (Fig.\ \ref{fig:1151-Model}), we have considered  two extreme cases: i) the gas is located in a disc at an average distance of 340 pc (half  the size of the radio source) and affected by the flux of one of the lobes, namely,\ $\sim$3.5~Jy, half of the total flux (solid line in Fig.\ \ref{fig:1151-tspin});
ii) the gas is located at a distance of 1 kpc, corresponding to the de-projected radius for a viewing angle of 20$^{\circ}$ (solid line).   
The plot of Fig.\ \ref{fig:1151-tspin} (obtained under the conservative assumption of the kinetic temperature $T_{\rm kin} =100$ K) shows that 
in either of the two cases (and for any realistic value of the \HI\ volume density), the \tspin\ is high, possibly up to a few thousand K (i.e.\ well above the conventional 100 K). 
High \tspin\ values ($\gtrsim 1000$ K) are consistent with those obtained, using the extinction estimates derived from the optical continuum modelling, in the case of the 2-Jy quasar PKS~1549--79, for which  a \tspin\ in the range $3000<$\tspin$<6000$ K was found (see \citealt{Holt06} for details). Also in this case, VLBI observations will be needed to derive more accurate parameters (\tspin\ and $n_{\rm \HI}$) for PKS~1151--34.

For \tspin = 1000 K (and a covering factor of unity) in the case of PKS~1151--34, we derived a column density of $N_{\rm \HI} = 1.3 \times 10^{21}$ cm$^{-2}$, which is not overly extreme and similar to what has been  obtained for other radio sources \citep{Morganti18}.
Thus, in this source, the UV luminosity is low, consistent with atomic neutral hydrogen being present close to the active SMBH. However, the \tspin\ of the gas is likely high, which explains the low optical depth of the \HI\ observed in this object. 

Finally, despite the confirmed presence of atomic hydrogen, no CO(1-0) has been detected with ALMA in this source \citep{Tadhunter24}.  
However, in agreement with the \HI\ detection, the Herschel observations from \citet{Bernhard22} and \citet{Dicken23} have shown  evidence of a substantial cool ISM ($\sim$$10^{7.5}$ \msun\ of dust). Thus, the non-detection of CO(1-0) could indicate that the cold molecular gas has extreme conditions, such as a high excitation temperature, resulting in the high J-transitions being more luminous compared to CO(1-0). This has been seen now in a number of cases (see e.g.\ \citealt{Oosterloo25} and refs therein, and Oosterloo et al.\ in prep). 

The presence of a rich medium in PKS~1151--34 is also suggested by the evidence of an interaction with a nearby companion galaxy affecting the PKS~1151--34 host. This interaction is clearly indicated by the second nucleus and distorted structures visible in the optical image of \cite{Ramos11}. Xshooter observations \citep{Santoro20} have confirmed that the second nucleus has a redshift close to that of the target   (see  Appendix \ref{sec:AppendixOpt1151} for further details). 

\subsubsection{H\,{\small I} in a young radio galaxy: PKS~1306--09}
\label{sec:1306}

Similar to the case of PKS~1151--34, PKS~1306--09 is  a young radio source. The radio continuum structure shows two lobes and a total projected size 
of about 2.4 kpc (see Fig.\ \ref{fig:TwoDetections2} \opm{right}).  PKS~1306--09 has a steep spectrum at high frequencies ($\alpha^{\rm 2.7~GHz}_{\rm 8~GHz}=-0.85$, \citealt{Tzioumis02}). Its size is consistent with those found for other CSS and with the lack of a flattening or peak at low frequencies (\citealt{ODea21}). As for  PKS~1151--34, these properties support the scenario in which the source is small because it is young, and not because of orientation or beaming effects.

In this source, we detected an absorption with extremely low optical depth ($\tau_{\rm peak}=0.0009$). 
Considering the strength of the source (5.4 Jy), this again illustrates the excellent spectral stability of MeerKAT.
As in the case  of PKS~1151--34, the absorption profile is broad  $W_{20} = 405$ \kms. However, in PKS~1306--09 the \HI\ profile appears highly blueshifted, by almost 500 \kms\  compared to the systemic redshift derived by \cite{Santoro20}. 
In addition, the width of the \HI\ profile is large ($\sigma = 110$ \kms) suggesting that the gas is not quiescent.
Thus, it is unlikely that the \HI\ is associated with a rotating disc. Instead, the \HI\ could be part of the gas outflow which has been  identified using X-shooter \citep{Santoro20} in the warm ionised gas. The broad \OIII$\lambda$5007 emission line tracing this outflow has been found to be extended in the same direction as the radio emission, suggesting a possible jet-driven outflow. Furthermore, the optical \OIII$\lambda$5007 emission-line profile is double-peaked and the blue peak has a velocity ($\sim$$-500$ \kms) that is close to that of the \HI\ absorption. 
Thus, the detection of \HI\ with similar blueshifted velocities at the location of the radio emission reinforces the scenario  that cold gas (\HI) is also associated with the outflow and that the outflow is being driven by the radio jets. This expands the number of cases, especially in young radio galaxies, where multi-phase outflows are observed, including gas in the cold phase (atomic neutral hydrogen and molecular; \citealt{Morganti18} and refs.\ therein; \citealt{Murthy24}).

We can estimate the mass outflow rate ($\dot{M}$) using  the expression from \citet{Heckman02}, expressed as
\begin{equation}
   \dot{M} = 30 \frac{\Omega}{4\pi} \times \frac{r_*}{\rm 1\,kpc} \times \frac{N_{\HI}}{10^{21} \rm cm^{-2}} \times \frac{v}{300\, \rm km s^{-1}}~ M_\odot~\rm year^{-1}
,\end{equation}
where $\Omega$ is the solid angle  the outflow subtends (which we assume to be   $\pi$), $r_*$ is the radius of the outflow, assumed to be 1.2 kpc (half of the source extent), $N_{\HI}$ is the column density of $0.4 \times 10^{21}$ cm$^{-2}$ estimated with the conservative assumption of \tspin=1000 K and assuming the \HI\ covers the entire source, $c_{\rm f} =1$.
This results in a mass outflow rate $\dot{M} = 2.5$ \msunyr. Although not extreme, this is consistent with what has been found  for \HI\ outflows in other radio AGNs (see Table 1 in \citealt{Morganti18}). Interestingly, this is also consistent with the outflow of warm ionised gas described in \cite{Santoro20}, where two approaches were used to estimate the mass outflow rate of the warm gas, resulting in lower limits between 0.4 and 1.4 \msunyr. 

As for PKS~1151--34,  no UV detection from GALEX is available for PKS~1306--09. Therefore, the UV luminosity was estimated from the extrapolation of the X-shooter fluxes in the same way as for PKS~1151--34 (see  Appendix \ref{sec:AppendixUV}). The resulting UV luminosity is low, $L_{\rm UV} = 8\times 10^{21}$ \WHz\ at 1216\AA, again below the limit proposed by \cite{Curran08} and suggesting conditions favourable for \HI\ to be detected.  
However, as in the case of PKS~1151--34, the high radio flux of the small radio source, together with the likely presence of jet-ISM interactions, suggest that in this source the gas could have a high \tspin\ (i.e.\ $\gtrsim 1000$ K), which  could be the reason for the low optical depth observed.

\subsection{Local intervening case }
\label{sec:DetectionLocalIntervening}

In one object, PKS~0405--12, the spectrum reveals  \HI\ absorption at the location of the southern lobe (see Fig.\ \ref{fig:0405}). 
The absorption is blueshifted by $\sim$1100 \kms\ with respect to the systemic velocity of  PKS~0405--12 ($z = 0.574$). The derived column density is $0.8 \times 10^{18}$ \tspin/$c_{\rm f}$. The offset in velocity and the off-centre location indicate a situation different from that in the other two sources.
The overlay of the radio continuum on the optical image (Fig.\ \ref{fig:0405} left) shows that a companion galaxy is located, in projection, about 100 kpc from PKS~0405--12, against the southern radio lobe where the absorption was detected. 
Thus, the absorption appears to be caused by this galaxy, which was identified by \citet{Marr85} as a large spiral galaxy\footnote{Coordinates RA: 04$^{\rm h}$ 07$^{\rm m}$ 48\fs2, Dec: --12\degr 11\arcmin 48\farcs9 (J2000)} with strong \OII\ emission lines at a redshift of $z=0.5678$. 
PKS~0405--12 is located in a group or cluster environment and this intervening galaxy appears to be part of this local environment. Thus, we consider the \HI\ \opm{detected  to} be an example of 'local intervening absorption'.

Such off-nucleus local-intervening  \HI\ absorptions are not very common, but a few cases are known. 
In two  of these, the  absorption appears to be associated with a giant gas disc not located close to the nucleus, but still associated with the host galaxy itself. Coma A \citep{Morganti02} is the clearest case, while PKS~1649--062 \citep{Curran11} is a possible other example. 

An instance that is even more similar to the case of PKS~0405--12 relates the \HI\ absorption  features associated with  galaxies located in the foreground of some radio sources. For example, \HI\ absorption from a small companion disc galaxy has been observed against the radio lobe of 3C~433 \citep{Murthy20}. The  column density measured in that case ($1 \times 10^{18}$ cm$^{-2}$ \tspin) is similar to the one detected in PKS~0405--12. Against the radio lobe of another source in the 2-Jy sample, PKS~0409--75, \cite{Mahony22} have found  absorption blueshifted by $\sim$3300 \kms\ compared to the optical redshift of the host galaxy and with a column density of $2.16 \times 10^{21}$ cm$^{-2}$, assuming a spin temperature of \tspin = 500 K and  covering factor of $c_{\rm f} = 0.3$. This would correspond to  $1.3 \times 10^{18}$ (\tspin/$c_{\rm f}$), comparable to what found for PKS~0405--12. 

The width of the absorption ($W_{20} = 125$ \kms) is narrower compared to those measured for the associated absorptions found in the present sample. 
This is, to first order, consistent with what is expected for intervening \HI\ absorption (see \citealt{Curran21} for a discussion).  However, the profile is relatively broad for an intervening absorber, with $W_0 \sim 160$ \kms, likely because the absorption is observed against an extended continuum source. This suggests that some of the spread in rotational velocity of the gas in the spiral galaxy is included. This can complicate the separation between associated and intervening absorbers based only on the width of the profile (see below). 

\begin{figure}
\centering
\includegraphics[width=0.95\linewidth]{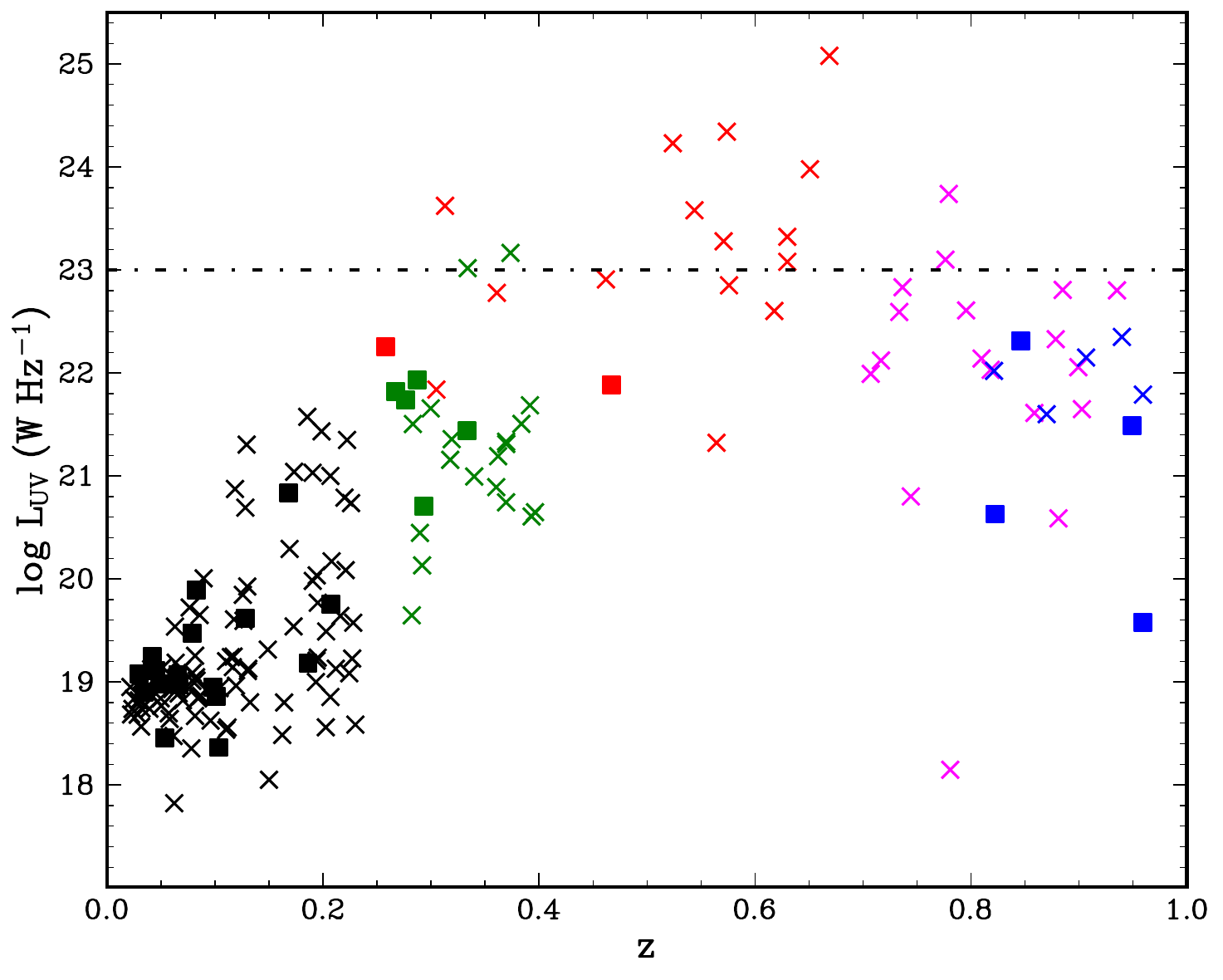}
\caption{\label{fig:Distribution-UV} UV luminosities for our sample (red points) together with some available in literature (black = \cite{Maccagni17}, green=\cite{Murthy21}, magenta = \cite{Murthy22}, and blue = \cite{Aditya19}, adapted from \cite{Murthy22}. Square points are detections, crosses represent the limits. The dashed line represents the  $L_{\rm UV} = 10^{23}$ \WHz\ threshold proposed by \citet[see text for details]{Curran08}.}
\end{figure}

\section{Origin and properties of the \HI\ absorbers}
\label{sec:Discussion}

In this work, we present observations searching for \HI\ absorption in a sample of powerful radio sources at intermediate redshift ($0.25<z<0.7$). The observations took advantage of the capabilities and performance (broad and stable bandpass, limited RFI, etc.) of the MeerKAT radio telescope. Because of this, given the brightness of the sources of the sample, we have been able to reach extremely low optical depth ($\tau_{\rm peak} \ll 0.01$). 
Some of our findings are similar to what could be expected based on  low-$z$ studies and we find no indication of major changes in the \HI\ properties at these intermediate redshift.

For 15 objects, we obtained useful \HI\ spectra (i.e.\ free from RFI), with three \HI\ absorptions detected, two associated absorptions in young radio sources and one so-called local intervening system  against the lobe of one of the extended radio sources and associated with a foreground galaxy. Thus, despite the high sensitivity, we obtained a relatively low detection rate  of associated absorption (i.e $13\%\pm 7\%$), albeit with a large uncertainty. 
The low detection rate is not completely surprising given that the sample includes mostly core dominated sources with broad optical emission lines, typical of quasar hosts. According to the unified schemes, dust and gas absorption are  mostly due to the circumnuclear torus and disc, which (in the case of these AGN types) are oriented close to face-on. Therefore, these AGNs are expected to be mostly unobscured. In this respect, the low detection rate can be attributed to the geometry of the absorber.

However, as described in the introduction, other effects can influence the detection rate, and one of the important parameters is the UV luminosity.  
Figure \ref{fig:Distribution-UV} shows the distribution of the UV luminosities derived as described in Appendix \ref{sec:AppendixUV}. About half of the sample has a luminosity above  $L_{\rm UV} = 10^{23}$ \WHz, namely, the threshold proposed by \citet{Curran08} as the limit above which  \HI\ can be ionised by the strong UV radiation. In agreement with this,  we have no \HI\ detections among those sources.  However, below this limit, associated \HI\ absorption has been detected for only two sources. 
Therefore, although it is difficult to completely distinguish between orientation and UV luminosity playing an important role  in whether the \HI\ absorption might be detected, our results  suggest that the UV luminosity is not the only parameter influencing the \HI\ detection and that the orientation is also an important parameter.

The optical depths of the \HI\  we observe are very low. Such low values could be reached because the target sources are strong and we obtained sensitive and high-quality observations.  In the case of our sources, the low optical depth could be the result of the gas having a high \tspin\ due to the combination of the gas being located  in the vicinity of the  active black hole and the strong radio flux. We show this for the case of one of the detections (PKS~1151--34). Thus, our results suggest that in general, the detection rate of \HI\ absorption should be interpreted with care, because it is strongly dependent on the properties of the studied sample and the depth of the observations. Well characterised samples and observations reaching low optical depth are needed to get a complete inventory of the presence of \HI.

Notably, we find that both detections of associated \HI\ are in young radio sources (i.e.\ two of the four CSS and peaked sources in our sample were detected). This confirms the trend seen both at low and high redshifts (see Sect.\ \ref{sec:Introduction}) and we see that even at intermediate redshift, they represent the most frequently detected sources in \HI. This trend is also starting to emerge from  blind \HI\ absorption surveys such as FLASH \citep{Yoon25}. 

The richness of the ISM observed in these sources is likely to be an important factor in their \HI\ detectability. Based on  the detected level of star formation activity and the amount of dust and cold molecular gas (see \citealt{ODea21,Dicken23,Tadhunter24} and refs therein), there is now considerable evidence that CSS and GPS have a richer ISM than their more extended counterparts. Interestingly, the presence of gas and the likely presence of interactions between jet and ISM, could, in fact, boost their radio emission and explain the large number of such sources detected in radio surveys (see \citealt{Morganti11} for details).
In addition to the gas, the fact that the size of the radio source matches that of the distribution of the gas in the circumnuclear regions could further help in boosting the detectability of these sources in \HI.  

In the context of the type of bias introduced by the selection of the sources in the present sample,  the case of the one broad-line, quasar-like object in our sample detected in \HI\ (PKS~1151--34) is interesting. On the basis of the orientation-based unified schemes for radio AGNs, 
we would expect the nucleus of this broad-line object to be relatively unobscured. Nevertheless \HI\ absorption has indeed been detected. 
For this quasar, which is classified as a young radio source, we suggest (based on the \HI\ profile, which is centred on the systemic velocity but highly asymmetric with the blueshifted peak much deeper) that the gas is distributed in a disc or ring with the radio source (two lobes with a projected size of 680 pc) still embedded in it (see Fig.\ \ref{fig:1151-Model}  for the proposed scenario). Thus, the nucleus may be partially obscured by this structure. Indeed, this is what is suggested by the presence of stellar absorption lines in its nuclear optical spectrum and the relatively red colours \citep{Santoro20}, indicating that a more complex situation than the standard circumnuclear torus and disc predicted from the unified scheme can be at work.

The \tspin\ in PKS~1151--34 is likely to be high, around 4000 K. This could explain the low optical depth of the detected gas ($\tau_{\rm peak} = 0.004$). If these conditions are also present in other sources, they could, as mentioned above,  affect the  detection fraction of  in observations of limited sensitivity.

In the second associated detection, PKS~1306--09, the \HI\ is found to be significantly blueshifted ($\sim 500$ \kms) compared to the systemic velocity.  Also, in this object, the \HI\ profile is broad (with velocity dispersion around 100 \kms) and has a low optical depth ($\tau_{\rm peak} = 0.0009$). This is another case that would be missed by shallow surveys. The large blueshifted velocity and the agreement found for the warm ionised gas (see Sect.\ \ref{sec:1306} and \citealt{Santoro20}) suggest that the \HI\ is part of a fast outflow, likely driven by the jet-ISM interaction. 
Although we cannot pinpoint the location of this blueshifted component (due to the limited resolution of our observations), we know that the lobes of the radio source are about 2 kpc apart, suggesting that the \HI\ ( if it is indeed interacting with the radio plasma) is on those scales. 
This finding expands on the number of young (or restarted) radio AGNs, where outflows of cold gas (atomic neutral hydrogen and cold molecular gas) are found. This also confirms the predictions of  simulations of the impact of the initial phase of the young radio jet on the surrounding medium, which is traced by the presence of gas outflows and large velocity dispersion of the gas due to the creation of a cocoon of shocked gas around the jet (\citealt{Mukherjee18} and refs therein). 

In contrast to the two associated absorption detections, a third absorption detection originates in an intervening companion galaxy located in front of the radio lobe of PKS~0405--12. The galaxy is likely part of the galaxy group surrounding the central quasar (with a difference in velocity of $\sim$1100 \kms); therefore, we have defined it as a local intervening case.  Interestingly, this absorption has a narrower profile compared to the associated ones.  
\cite{Curran21} has proposed to use the width of the \HI\ profile as a criterion to separate associated from intervening absorption in  large blind \HI\ absorption surveys. 
However, according to \citet{Curran21}, only for FWZI $\leq 50$ \kms\ there is a $\sim$90 per cent probability of the absorber being intervening. For cases of local intervening absorption such as PKS~0405--12, the broader profile (FWZI $\sim 160$ \kms) is likely due to the inclusion of some of the rotation of the gas in the intervening galaxy because the absoprtion is observed against an extended continuum background. This is relevant because it indicates that for blind surveys, it will be difficult to classify a group of \HI\ profiles  as  'associated' or 'intervening', unless the radio data are characterised by a good spatial resolution and/or complementary optical data are available to identify these situations. 

\section{Conclusions}
\label{sec:Conclusions}

Thanks to the well-characterised sources and deep radio observations, the sample of radio AGN targets of this study  has shown a variety of \HI\ absorption properties, similarly to what has been found for low-$z$ objects, such as the low-$z$ part of the 2~Jy sample presented in \citet{Morganti01}.
The sample presented here is dominated by broad-line sources where, according to the unified schemes,  the nucleus should be unobscured. Indeed, we do have a relatively low detection rate ($13\%\pm 7\%$), with the majority of the sources are not detected, even if the UV luminosity is below the  limit of $10^{23}$ \WHz. 

However, young (CSS and peaked) radio sources, also at intermediate redshifts, are confirmed to be  more likely to be detected, even in the case where they are quasars. It is, however, important to note that the two detections have extreme low optical depths ($\tau_{\rm peak}<0.005$) and could only be found  thanks to the depth and stability of the MeerKAT observations presented here.
The higher occurrence of \HI\ in CSS and peaked sources is often considered to be related to the richness of their ISM and the results from our study further confirm this. This highlights the need for ancillary data to complement the \HI\ observations.  This is important for the blind surveys such as FLASH \citep{Yoon25}, which are now expanding the view of atomic neutral hydrogen in AGNs to higher redshifts. 

In general, the results (i.e. detection rate and properties of the absorption) for the present study, which is focussed on well characterised (at multi-wavelengths) target sources, can guide the interpretation of the results from untargeted surveys. 
Follow-up \HI\ high-resolution (VLBI) observations are needed to confirm the scenarios we propose about the distribution of the absorber in PKS~1151--34 and the presence of an \HI\ outflow in PKS~1306--09. The VLBI capability of SKA will be an extremely valuable addition to trace the \ HI\ on parsec scales.
At the same time, observations of multiple transitions of the molecular gas observed in emission  (e.g.\ CO) would help to elucidate more information on the full distribution and physical conditions of the gas, overcoming the limitations of absorption studies. 

The sample presented here is still biased towards core-dominated objects and against extended, FRII-like, sources. Expanding to the latter group of objects will be the next  step to test the presence of circumnuclear \HI\ discs as main absorber in radio AGNs, thereby reinforcing the role of orientation in the detection rate of \HI\ absorption. However, such observations will prove to be time-consuming because the cores of these sources typically  exhibit low radio flux.
In general, the present study shows that the limitations of the studied samples, in terms of selecting the targets and depth of the observations, should be taken into account when interpreting the detection rate and the properties of the \HI\ gas in radio AGNs.

\begin{acknowledgements}
The MeerKAT telescope is operated by the South African Radio Astronomy Observatory, which is a facility of the National Research Foundation, an agency of the Department of Science and Innovation.  
This work used images from the DESI Legacy Imaging Surveys \citep{Dey19}. Their full acknowledgment
is at https://www.legacysurvey.org/acknowledgment/.

\end{acknowledgements}

\begin{appendix}

\section{UV luminosities} 
\label{sec:AppendixUV}  

The UV luminosities for the 2-Jy objects were obtained using different approaches depending on the data available. The luminosities were  corrected to 1216 \AA\ to make them consistent with \cite{Curran08}.
No correction has been made for Galactic dust extinction.

A group of objects (PKS~0403--13, PKS~0405--12, PKS~0637--75, PKS~1136--13, PKS1954-388, PKS~2203--18, PKS~2243--12 and PKS~2345--16) were observed and detected with GALEX. For these objects we extrapolated down to 1216 \AA\ assuming a power law and a spectral index derived from the NUV and FUV measurements. 

A second group (PKS~0252--71, PKS~1151--54, PKS~1306--09), which include the two \HI\ detected objects, is made of radio galaxies that were not detected  nor observed by GALEX, but with 
VLT Xshooter observations (all CSS sources) from \cite{Santoro20}.
In these cases, we extrapolated from the measured flux typically  from the Xshooter spectra at rest-frame wavelength $\sim$2200 - 2700 \AA) to 1216\AA\ using the \cite{Berk01} quasar SED. 
The correction factor is typically quite large (2.4), so these are very much upper limits because use of the quasar SED is likely to substantially over-estimate the true UV fluxes of these galaxies, for which the optical/UV light is likely to have a significant stellar component.

PKS1355--41 and PKS1510--08 were not observed by GALEX, but have been observed in the UV with HST/COS or IUE,  covering the redshifted wavelength of  Lyman$\alpha$.

Finally a group of four objects (PKS0159--11, PKS0842--75, PKS0859--25 and PKS1602+01) have no observations in the UV. 
For these, we have  taken the 3500\AA\ (rest) fluxes from \cite{Tadhunter93}  of the optical spectra and extrapolated to 1216\AA\ using the \cite{Berk01} quasar SED. For these cases, the correction factor is large  (4.2).

\section{Bandpass stability of the observations of PKS~1306--09} 
\label{sec:AppendixBP}  

\begin{figure}[h]
\centering
\includegraphics[width=0.91\linewidth]{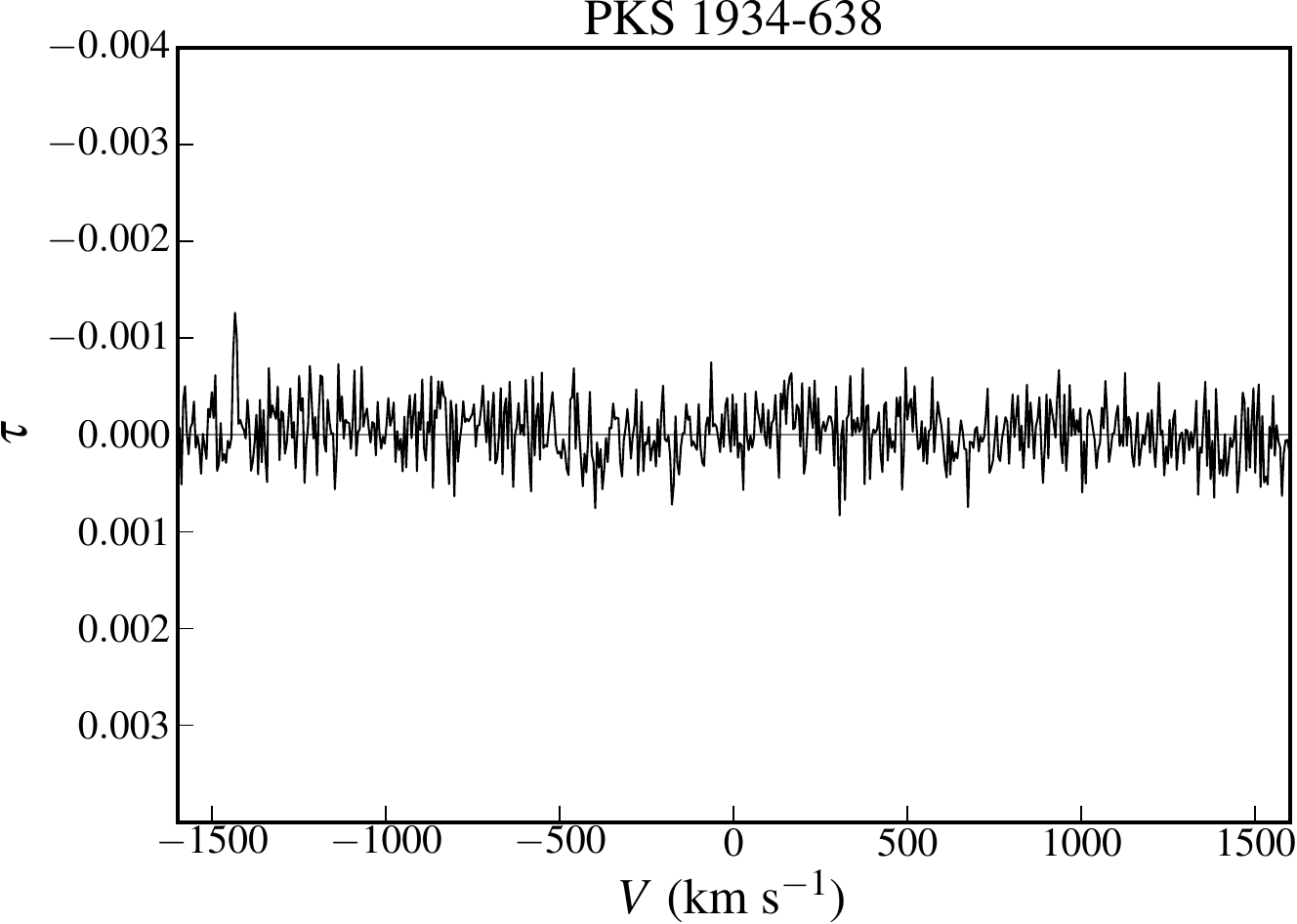} 
\caption{Spectrum of PKS 1934--63 calibrated using the bandpass derived from J0408--6545 observed several hours later.
\label{fig:bandpass} }
\end{figure}

To illustrate the good quality and stability of the bandpass calibration, we applied the bandpass calibration obtained using J0408--6545 which was observed at the end of the observing block used for PKS~1306--09, to the observation of PKS 1934--63 which was observed at the beginning of this observing block. The resulting spectrum (expressed in optical depth), is shown in Fig.\ \ref{fig:bandpass} and shows that this spectrum is noise dominated.

\section{Optical data of PKS~1151--34 and companion}
\label{sec:AppendixOpt1151}     

\begin{figure}[h]
\centering
\includegraphics[width=0.8\linewidth]{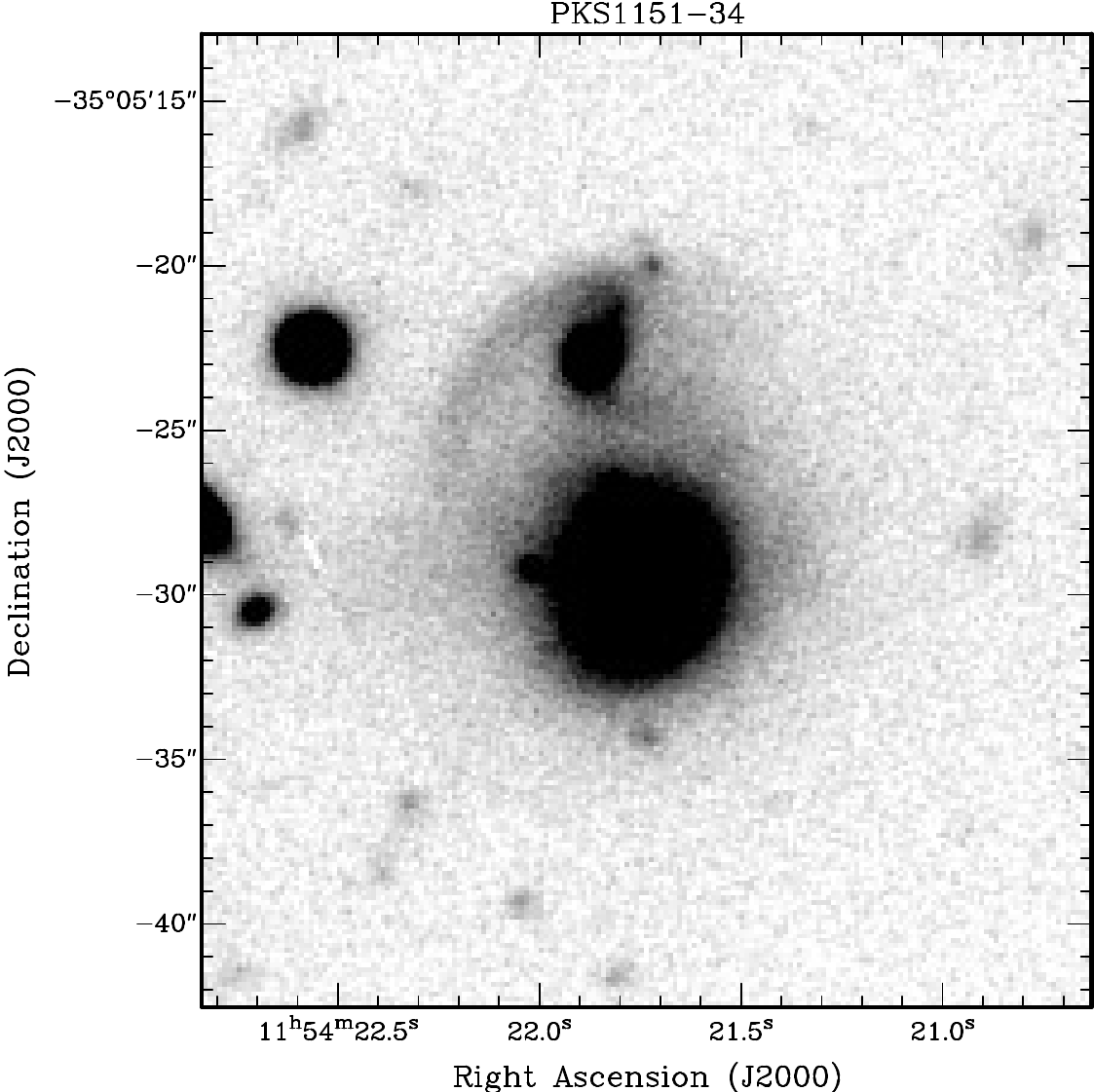} 
\caption{\label{fig:1151opt}  GMOS-S/Gemini optical broad-band image of PKS~1151--34 (taken from \citealt{Ramos11}).} 
\end{figure}

Figure \ref{fig:1151opt} shows the GMOS-S/Gemini optical broad-band image of PKS~1151--34 from \citet{Ramos11}. The presence of a companion SW of the host galaxy of PKS~1151--34 can be clearly seen. A tail appears to connect the two galaxies. To confirm that the two objects are indeed at similar distance, we use the X-Shooter spectrum published by \cite{Santoro20} which, interestingly, had the slit aligned with the nucleus of the companion. The spectrum shows clear absorption lines, so the redshift could be measured. From measurements of 4 absorption lines in this Xshooter spectrum the derived redshift is  $z_{\rm comp} = 0.25702 \pm 0.00003$. This means that we confirm the galaxy as a companion, blueshifted by --215 $\pm$ 25 \kms\ in the rest-frame of PKS~1151--34 and blueward of the strongest \HI\ absorption.

\section{Radio continuum images}
\label{sec:AppendixFigRadioCont}        

\begin{figure*}
   \centering
\includegraphics[width=5.4cm]{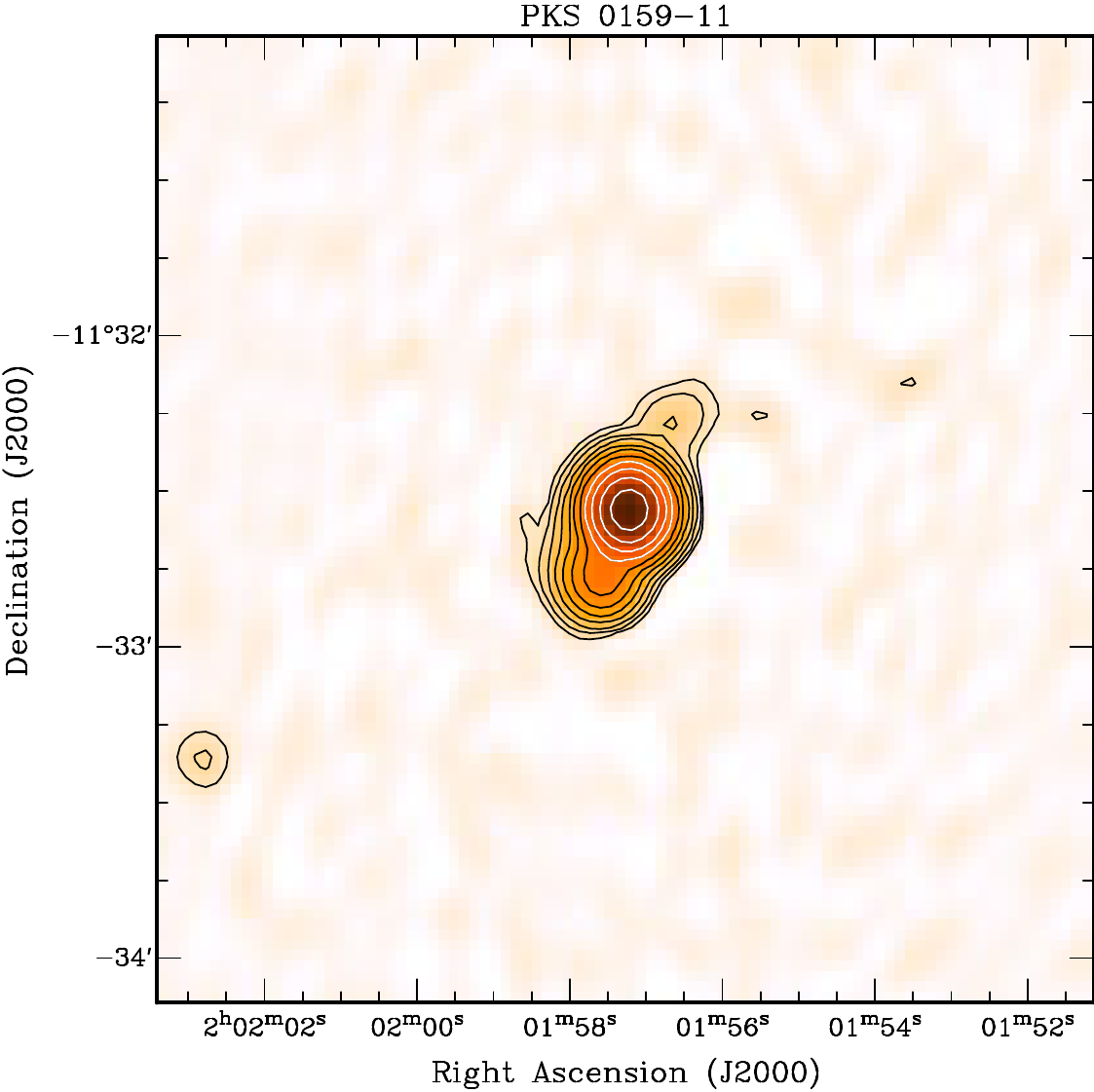} 
\includegraphics[width=5.4cm]{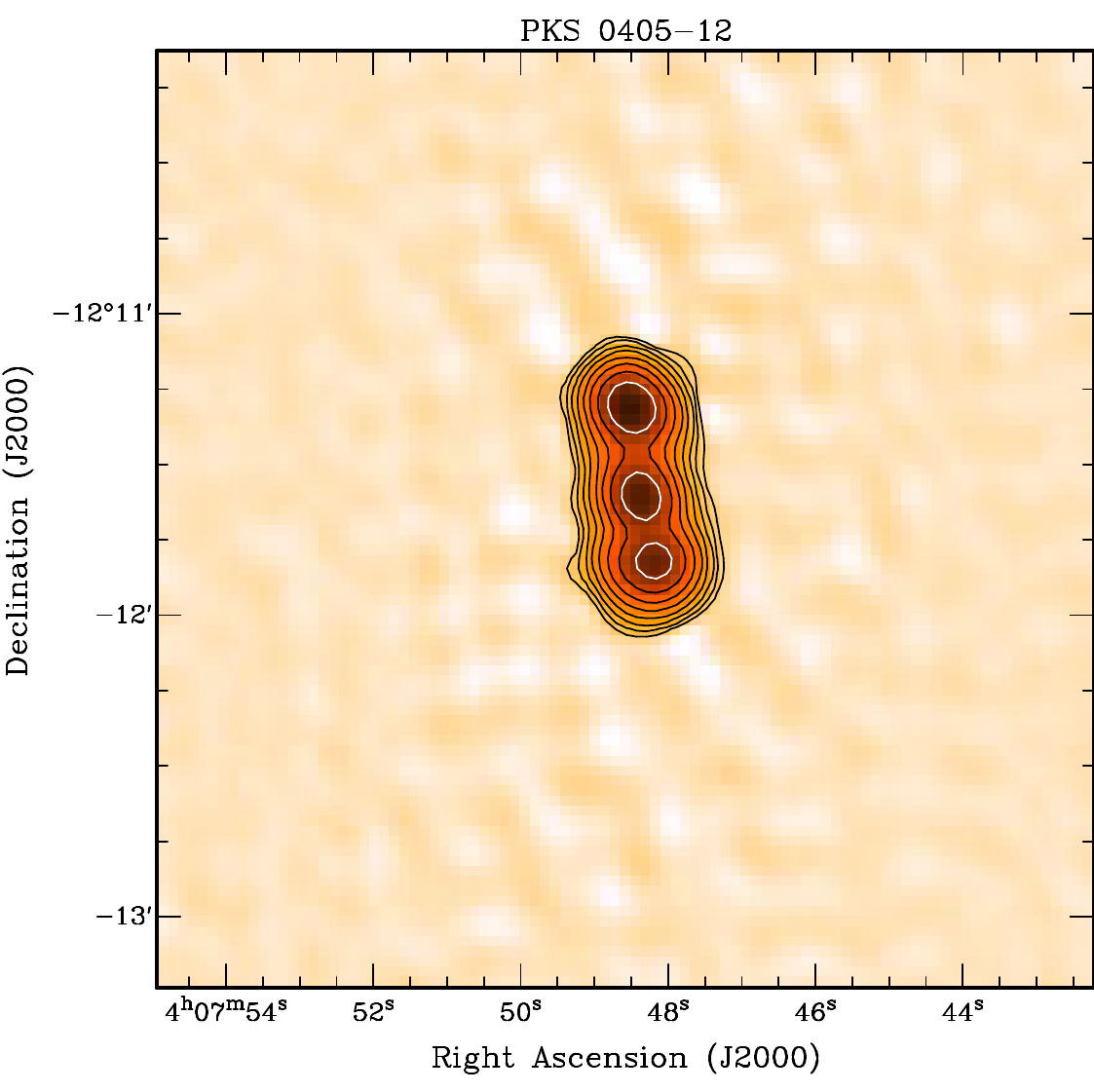} 
\includegraphics[width=5.4cm]{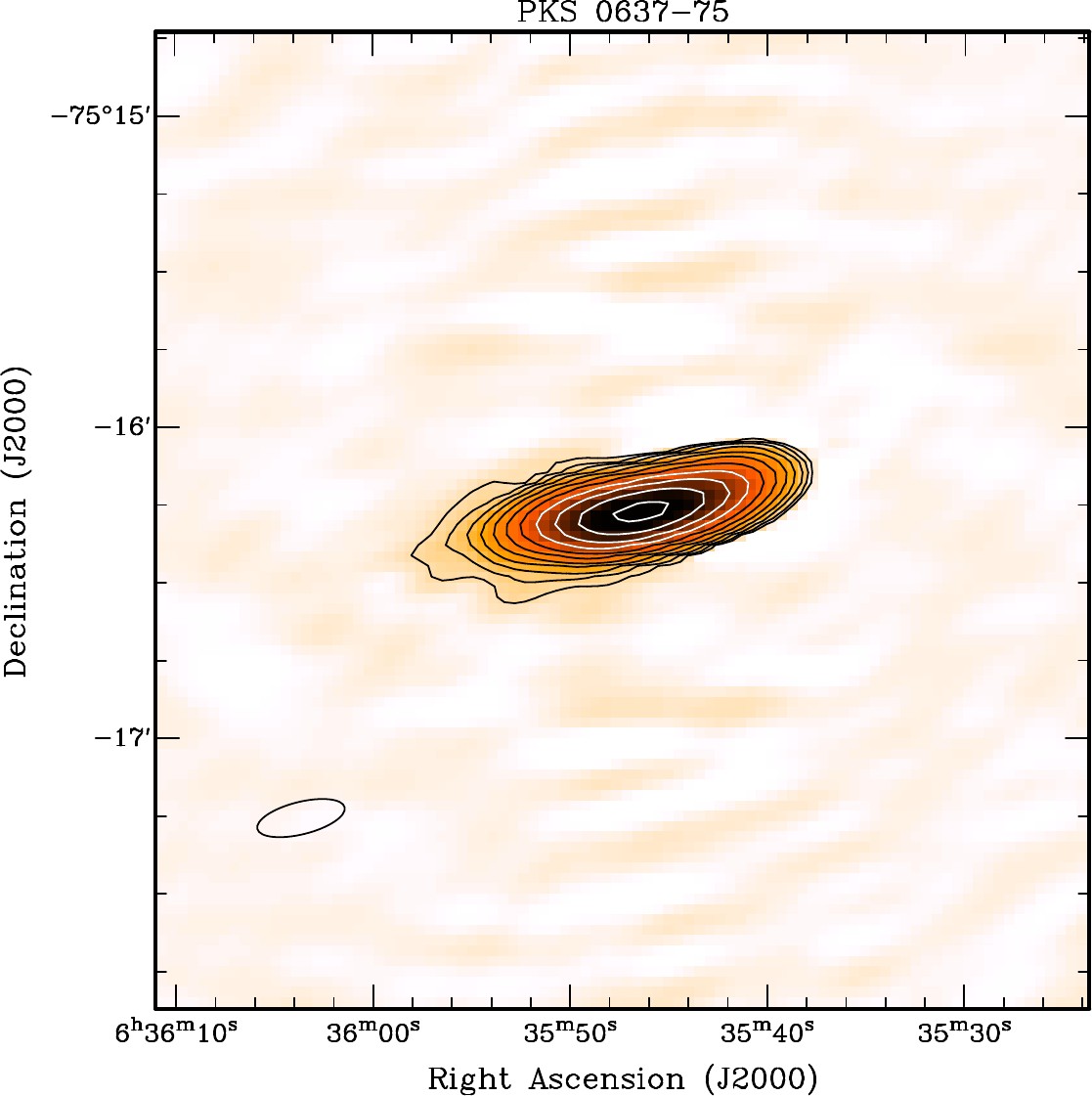} 
\includegraphics[width=5.4cm]{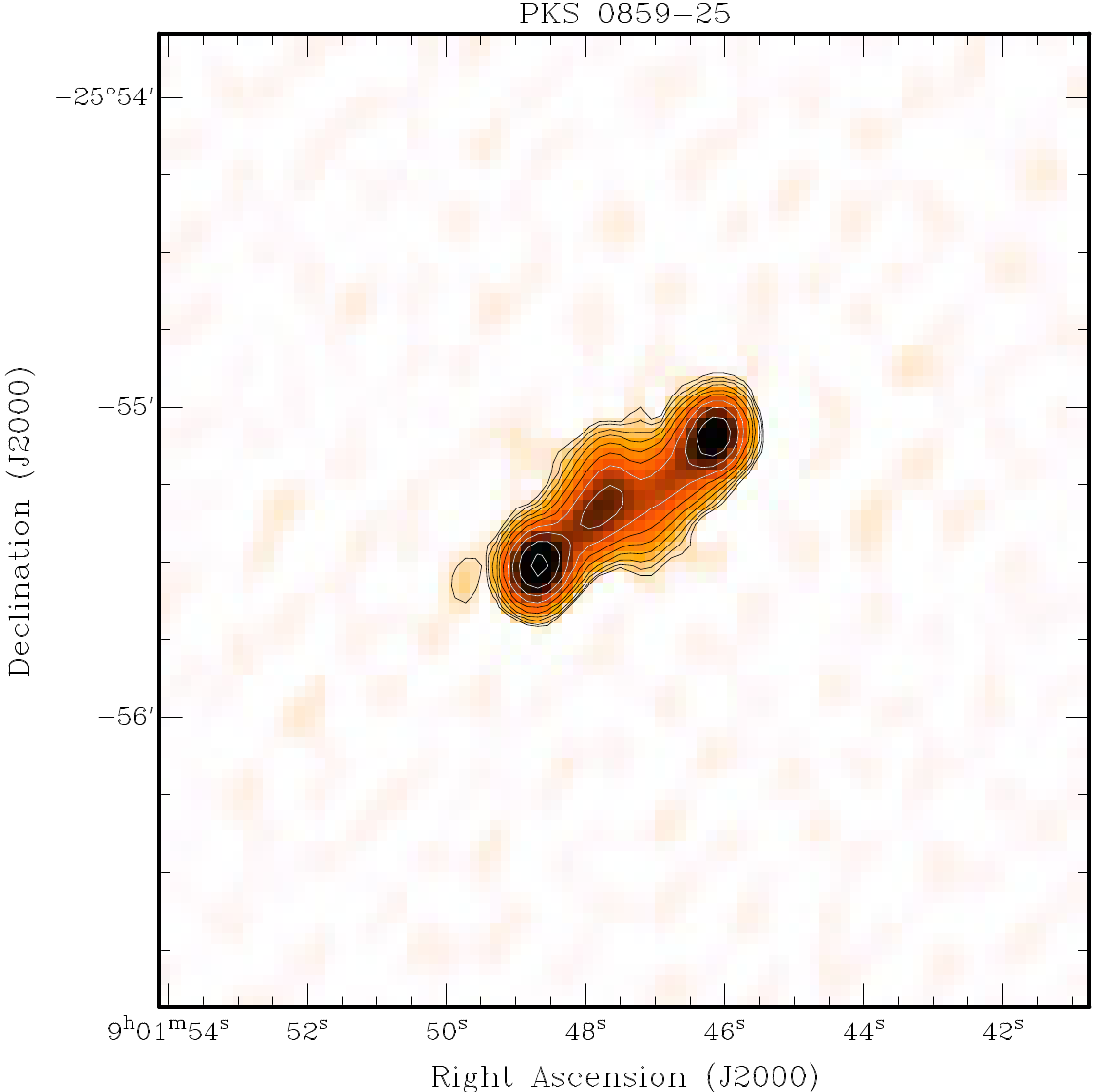} 
\includegraphics[width=5.4cm]{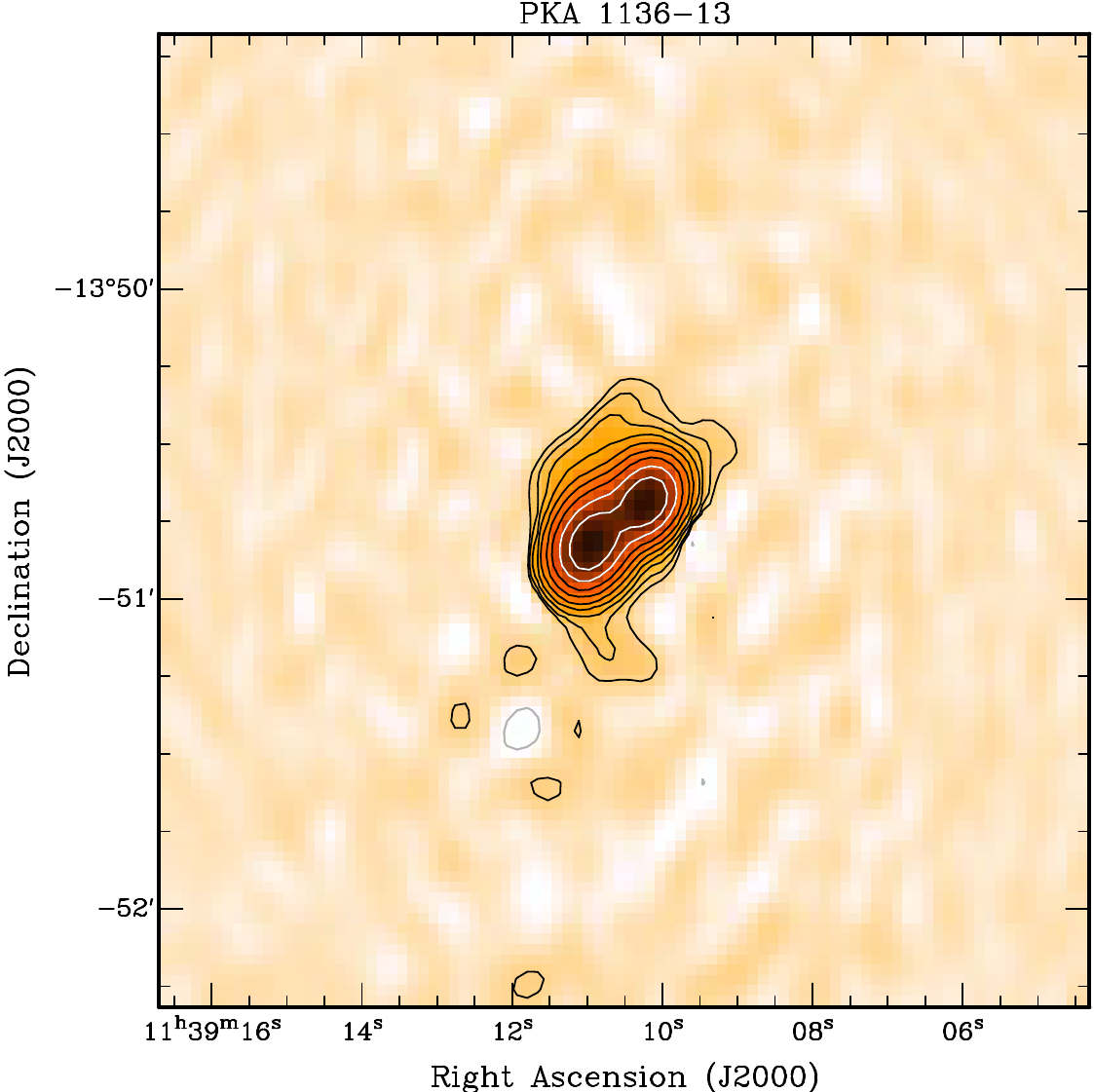} 
\includegraphics[width=5.4cm]{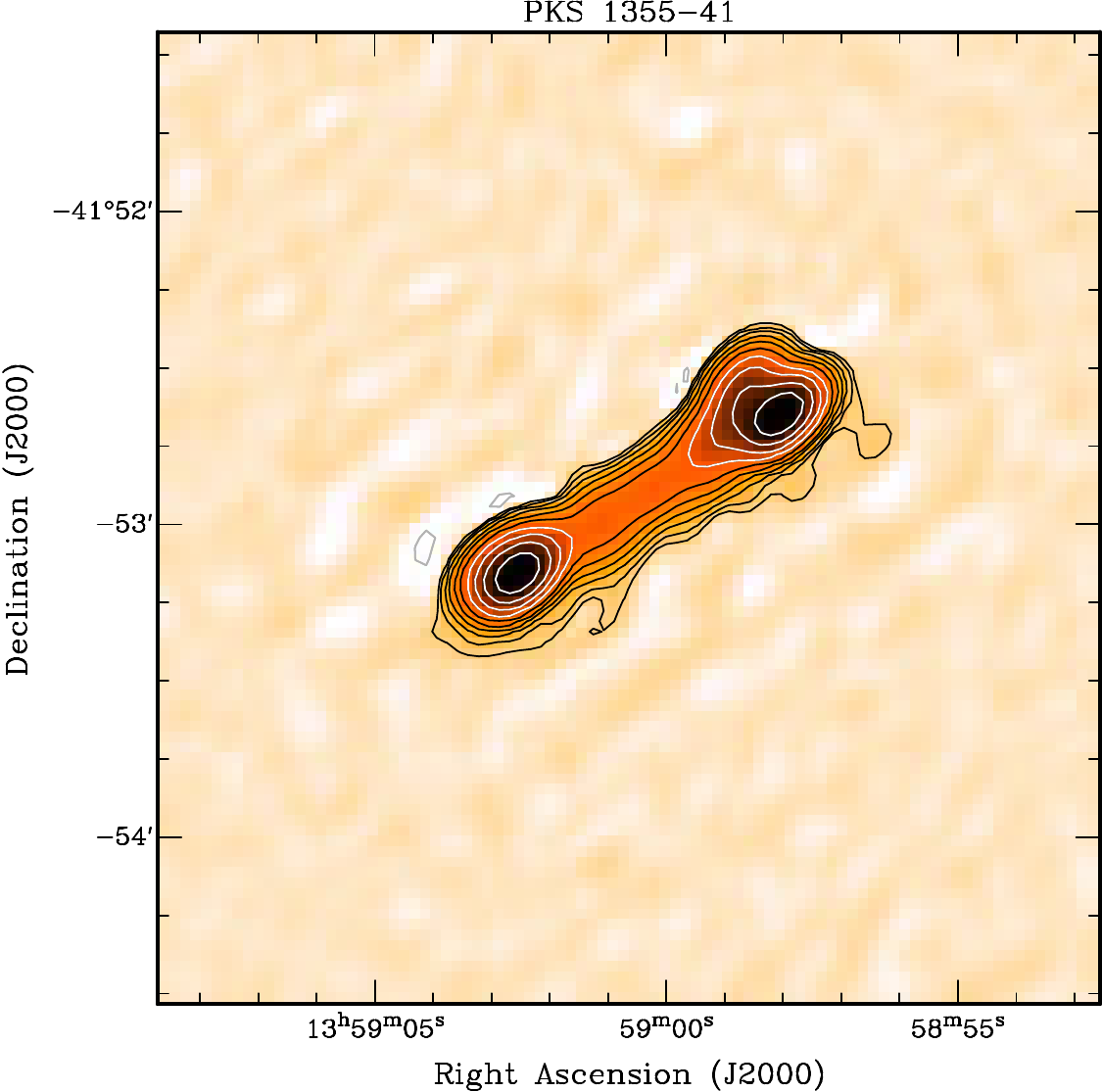} 
\includegraphics[width=5.4cm]{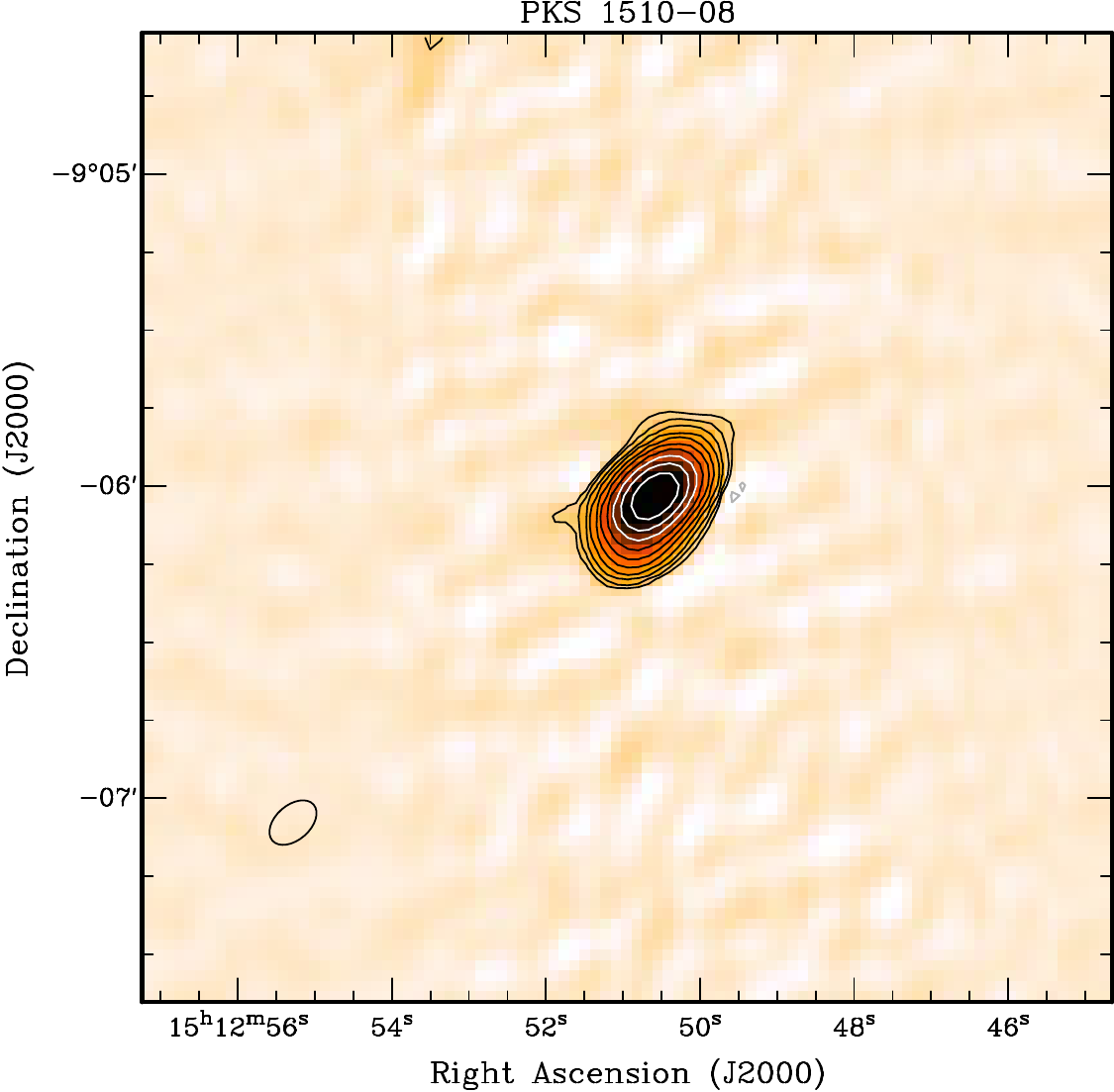} 
\includegraphics[width=5.4cm]{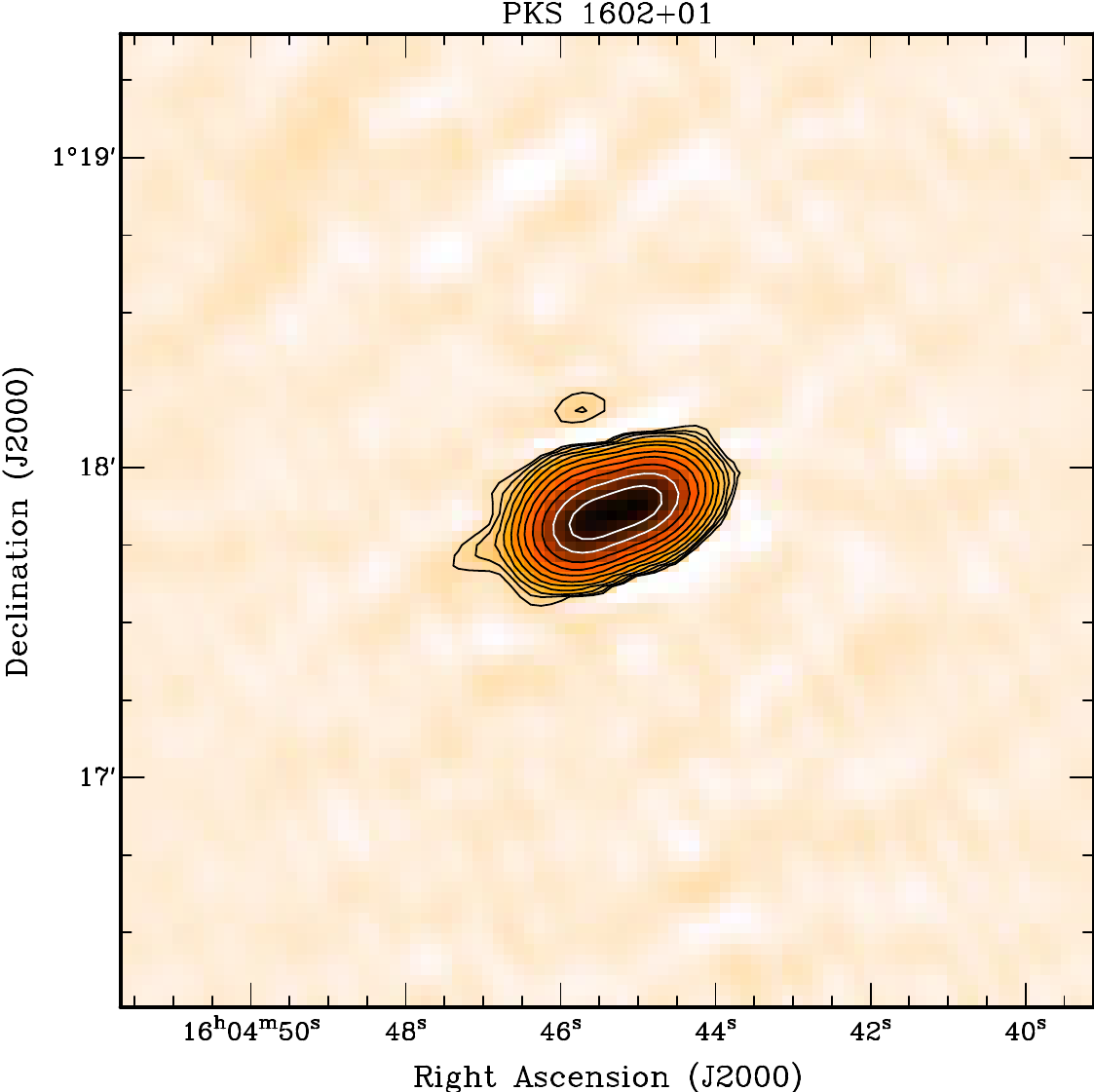} 
\includegraphics[width=5.4cm]{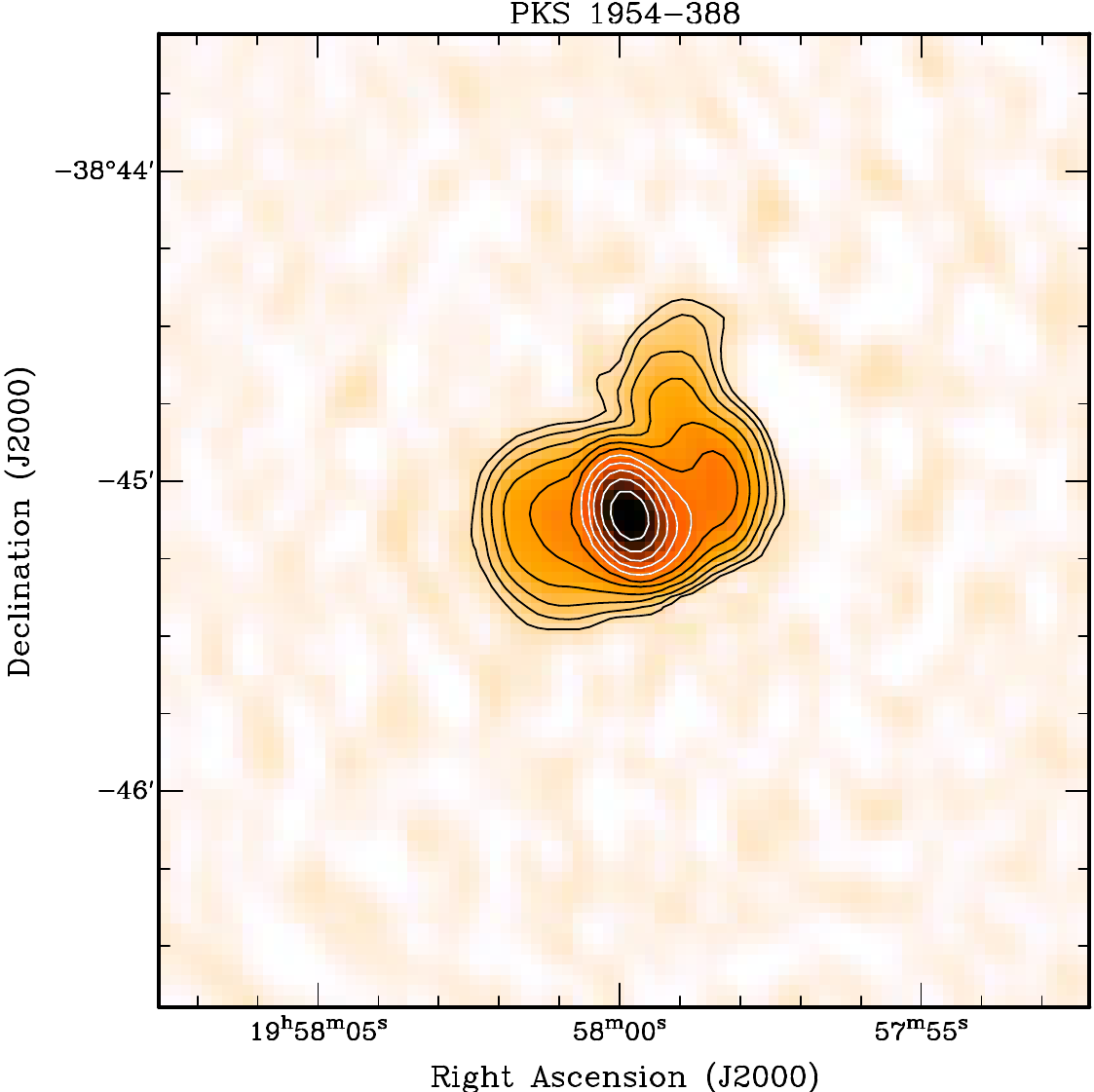} 
\includegraphics[width=5.4cm]{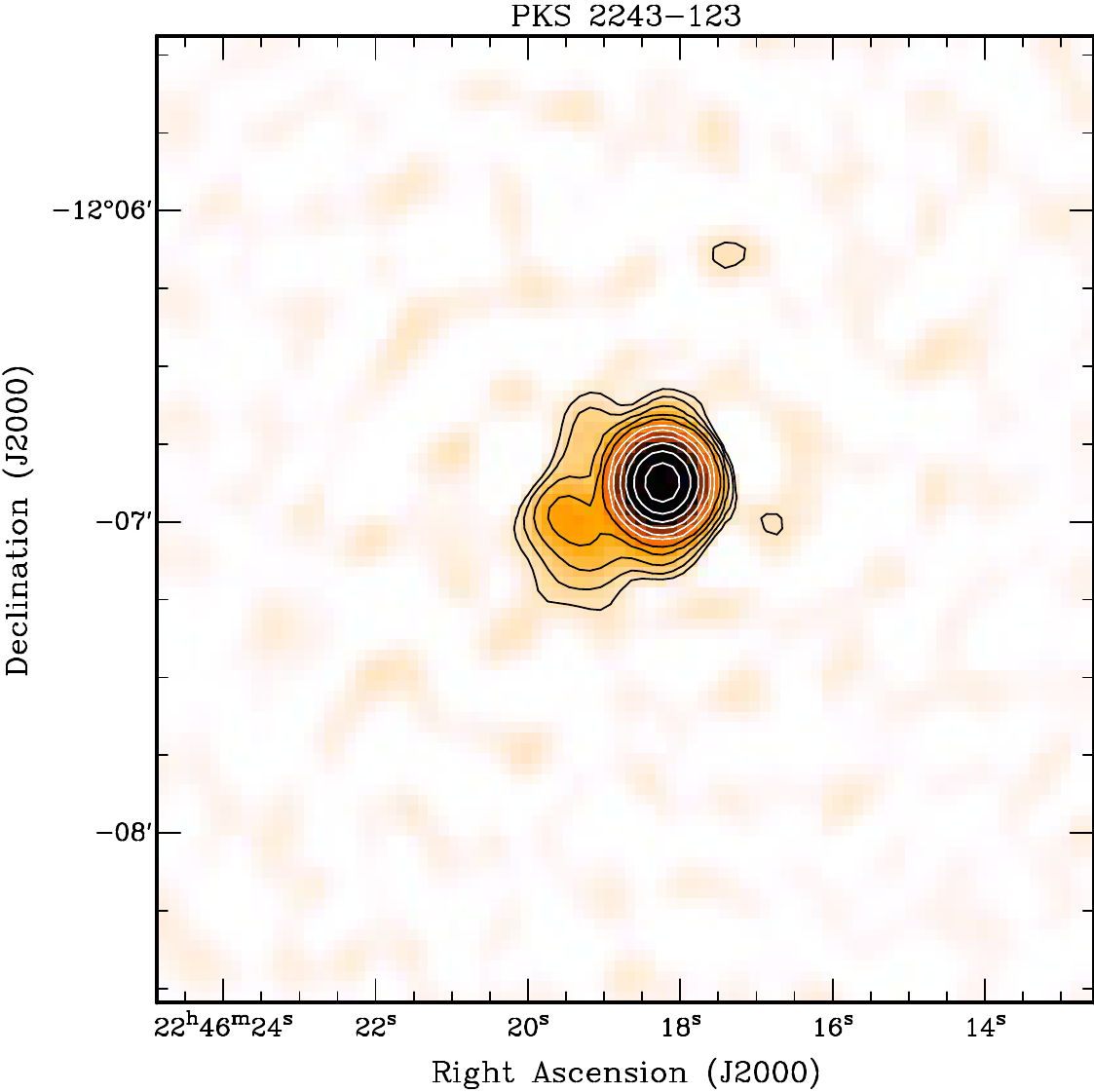} 
\caption{Images of the sources with extended radio continuum emission. PKS 0159--11.:Contour levels are 2, 4, 8, ... \mJybeam; PKS 0405--12: Contour levels are 5, 10, 20 ... \mJybeam; PKS 0637--75: Contour levels are 5, 10, 20, ... \mJybeam; PKS 0859--25: Contour levels are 4, 8, 16, 32, ... \mJybeam; PKS 1136--13: Contour levels 5, 10, 20, 40, ... \mJybeam ; PKS 1355--41: Contour levels are 2, 4, 8, 16, ... \mJybeam;  PKS 1510--08: Contour levels are 3, 6, 12, ... \mJybeam; PKS 1602+02: Contour levels are 2, 4, 8, ... \mJybeam; PKS 1954--388: Contour levels are 0.5, 1, 2, 4, ... \mJybeam; PKS 2243--1123: Contour levels are  1, 2, 4, ... \mJybeam..
}
\label{fig:imagesCont}
\end{figure*}

Figure \ref{fig:imagesCont} shows the radio continuum images of the extended sources. The parameters of the images are given in Table \ref{tab:obs}.

\end{appendix}

\end{document}